\documentclass[useAMS,usenatbib]{mn2e}
\usepackage{epsfig}
\usepackage{graphics}
\usepackage{afterpage}
\usepackage{times}
\usepackage{amsmath}
\usepackage{amssymb}

\newlength{\colwidth}
\newcommand{\ion}[2]{\hbox{#1\,{\sc #2}}}

\newcommand{\Msun}{{{\rm M}_\odot}}
\newcommand{\hMsun}{{h^{-1}\,{\rm M}_\odot}}

\newcommand{\apj}{ApJ}
\newcommand{\apjl}{ApJL}

\newcommand{\mnras}{MNRAS}
\newcommand{\aj}{AJ}
\newcommand{\apjs}{ApJS}
\newcommand{\nat}{{\it Nature}}

\newcommand{\pasp}{PASP}

\newcommand{\procspie}{SPIE}

\title[Halo mass from Ly$\alpha$ absorption profiles]{A measurement of galaxy halo mass from the surrounding \ion{H}{I} Ly$\alpha$ absorption} 
\author[O. Rakic et al.]{%
Olivera~Rakic,$^{1,2}$\thanks{E-mail: rakic@strw.leidenuniv.nl} 
Joop~Schaye,$^1$ Charles~C.~Steidel,$^3$ 
Craig Booth$^{4,5}$, \newauthor
Claudio Dalla Vecchia $^6$, 
and Gwen~C.~Rudie$^3$\\  
\\
$^{1}$ Leiden Observatory, Leiden University, P.O. Box 9513, 2300 RA Leiden,
  the Netherlands\\
$^{2}$  Max-Planck-Institut f\" ur Astronomie, K\" onigstuhl 17, 69117, Heidelberg, Germany\\
$^3$ California Institute of Technology, MS 249-17, Pasadena, CA 91125, USA\\
$^4$ Department of Astronomy \& Astrophysics, The University of Chicago, Chicago, IL 60637, USA\\
$^5$Kavli Institute for Cosmological Physics and Enrico Fermi Institute, The University of Chicago, Chicago, IL 60637, USA\\
$^6$Max Planck Institut f\" ur Extraterrestrische Physik, Giessenbachstra§e 1, 85748 Garching, Germany
}

\begin{document}

\pagerange{\pageref{firstpage}--\pageref{lastpage}} \pubyear{2012}

\maketitle

\label{firstpage}

\begin{abstract}
We measure the dark matter halo masses of $\langle z\rangle\approx2.36$ UV color-selected star-forming galaxies by matching the observed median \ion{H}{I} Ly$\alpha$ absorption around them, as observed in the spectra of background QSOs, to the absorption around haloes above a given mass in cosmological simulations. Focusing on transverse separations 0-2 pMpc and line of sight separations 154-616 $\rm km\, s^{-1}$, we find a minimum halo mass of log$_{10}M_{\rm min}/\Msun={11.6\pm0.2}$, which is in good agreement with published halo mass estimates from clustering analyses. We verified that the measured halo mass is insensitive to a change in the cosmological parameters (WMAP1 vs.\ WMAP3) and to the inclusion of strong AGN feedback. One unique strength of this method is that it can be used in narrow field galaxy-QSO surveys, i.e.\  $\approx30\times30$ arcseconds. In addition, we find that the observed anisotropy in the 2-D \ion{H}{I} Ly$\alpha$ absorption distribution on scales of 1.5-2 pMpc is consistent with being a consequence of large-scale gas infall into the potential wells occupied by galaxies.   
\end{abstract}


\def\figb{
\begin{figure*}
\centering
\resizebox{0.25\textwidth}{!}{\includegraphics*[trim = 0mm 0mm 85mm 0mm, clip]{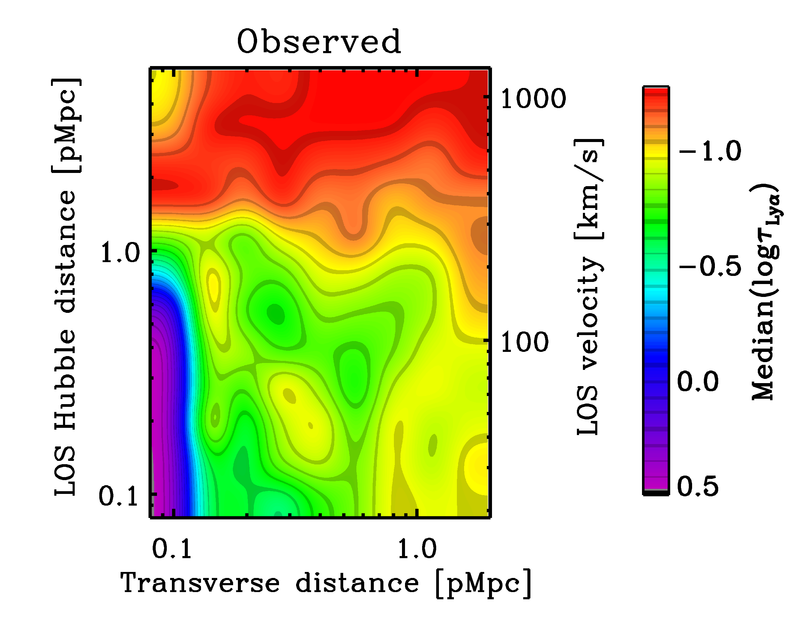}}\resizebox{0.21\textwidth}{!}{\includegraphics*[trim = 30mm 0mm 75mm 0mm, clip]{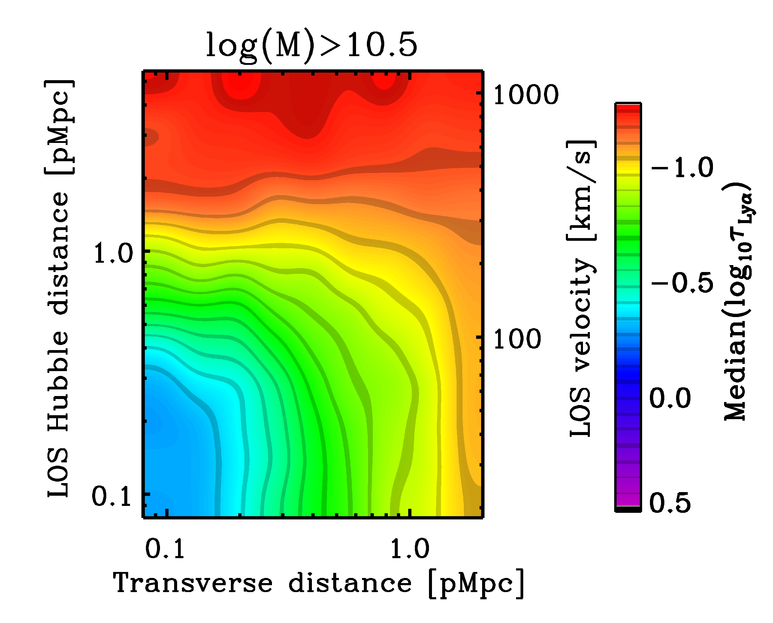}}\resizebox{0.21\textwidth}{!}{\includegraphics*[trim = 30mm 0mm 75mm 0mm, clip]{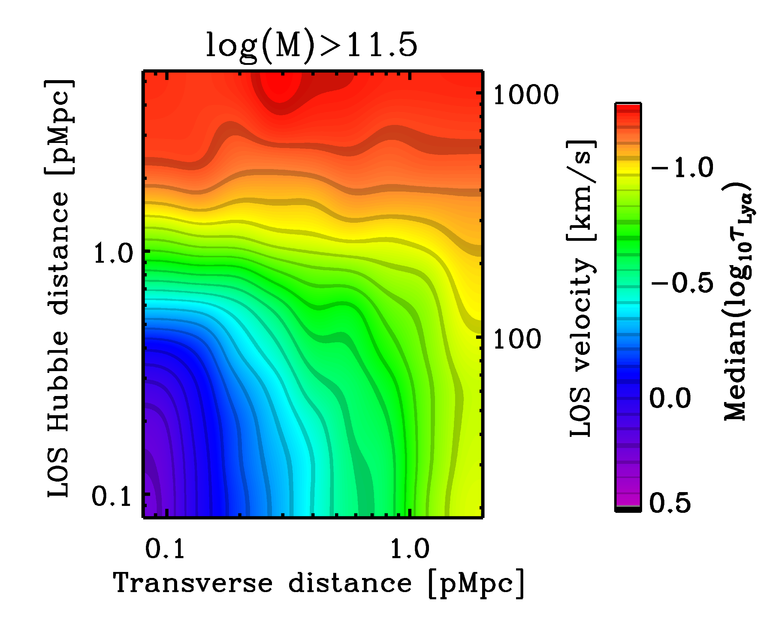}}\resizebox{0.303\textwidth}{!}{\includegraphics*[trim = 30mm 0mm 0mm 0mm, clip]{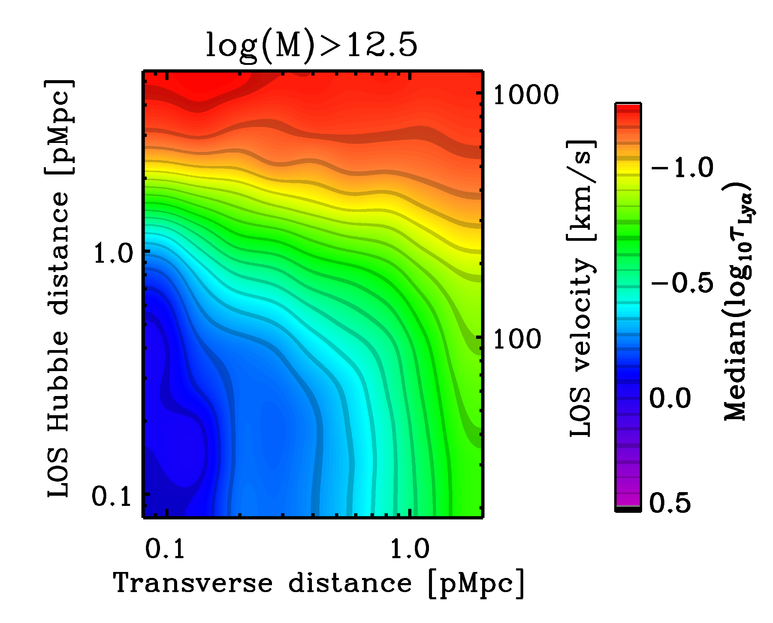}}\caption{Median 2-D distribution of \ion{H}{I} Ly$\alpha$ absorption around galaxies, from left to right, as observed by \citet[][their Figure 5]{Rakic2012}, and in the \emph{REF} simulations near haloes with $M_{\rm min}>10^{10.5}$ (using the 50 $h^{-1}$ cMpc box), $10^{11.5}$ (using the 50 $h^{-1}$ cMpc box) and $10^{12.5}\Msun$ (using the 100 $h^{-1}$ cMpc box) at $z=2.25$. Redshift errors (as described in the text) were included in the simulated maps. In comparison with the observed absorption map,  the second panel shows too little absorption at large impact parameters, while the fourth panel shows too much absorption in those regions. \label{2D_masses}}  
 \end{figure*}
}

\def\figba{
\begin{figure}
\centering
\resizebox{0.35\textwidth}{!}{\includegraphics*{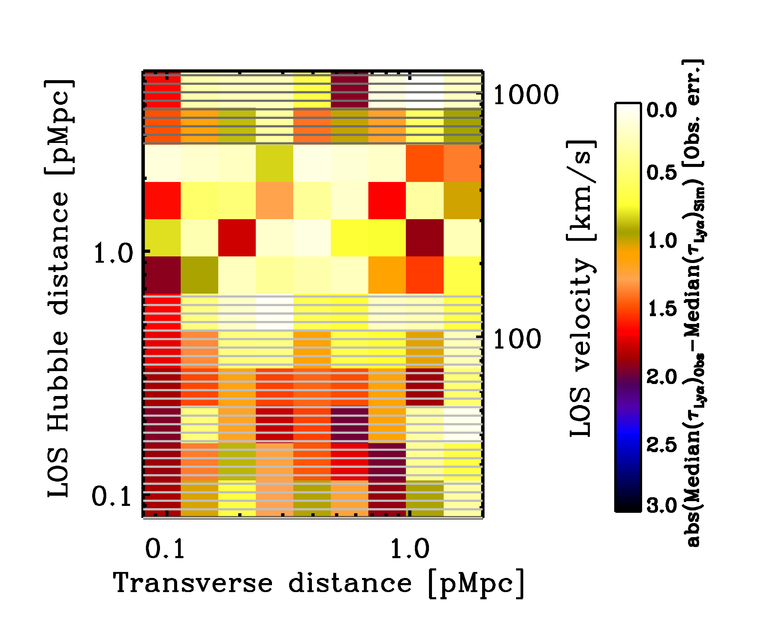}}
\caption{The absolute difference between the observed map and the closest matching \emph{REF} simulated map, for $M_{\rm min}>10^{10.6}\Msun$ using the 100 $h^{-1}$ cMpc simulation box, in units of the observational error. The shaded regions were not used in the comparison, due to uncertain baryonic physics and redshift errors at velocity separations  $<154\, \rm km\, s^{-1}$, and because the absorption signal is consistent with noise at separations larger than 616 $\rm km\, s^{-1}$ (see text for more details). \label{2D_difference}}  
 \end{figure}
}

\def\figbb{
\begin{figure*}
\centering
\resizebox{0.202\textwidth}{!}{\includegraphics*[trim = 0mm 0mm 85mm 0mm, clip]{F2.png}}\resizebox{0.17\textwidth}{!}{\includegraphics*[trim = 30mm 0mm 75mm 0mm, clip]{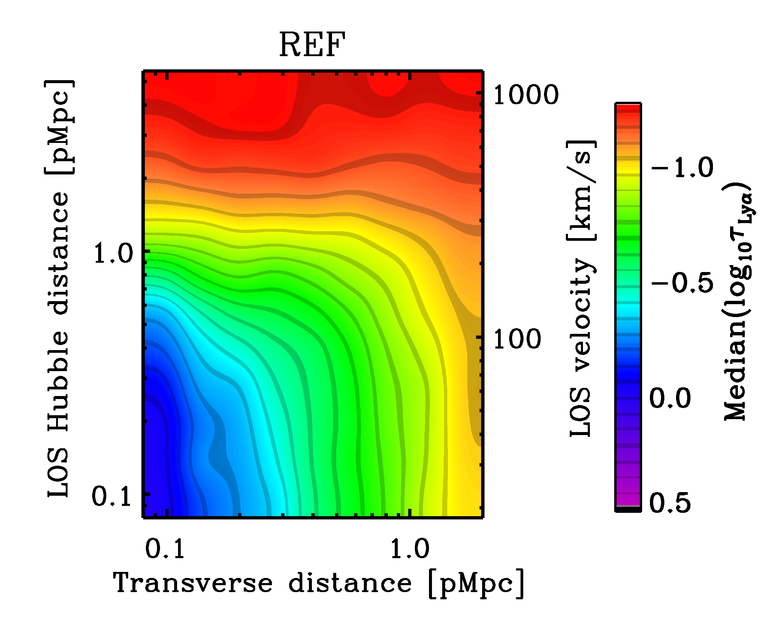}}\resizebox{0.17\textwidth}{!}{\includegraphics*[trim = 30mm 0mm 75mm 0mm, clip]{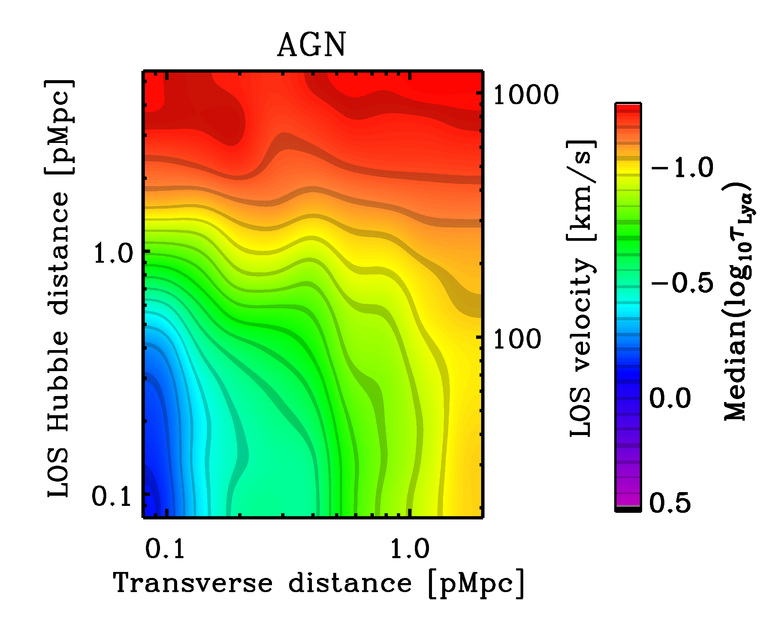}}\resizebox{0.17\textwidth}{!}{\includegraphics*[trim = 30mm 0mm 75mm 0mm, clip]{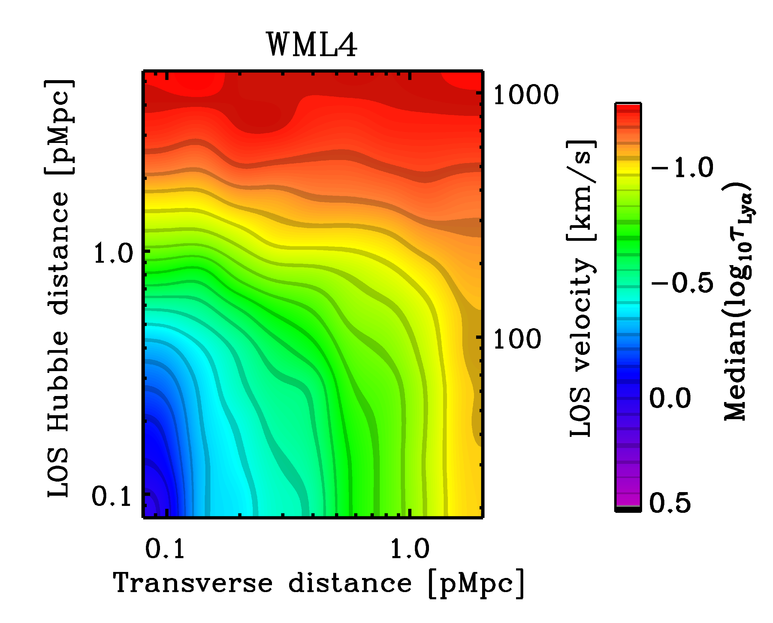}}\resizebox{0.242\textwidth}{!}{\includegraphics*[trim = 30mm 0mm 0mm 0mm, clip]{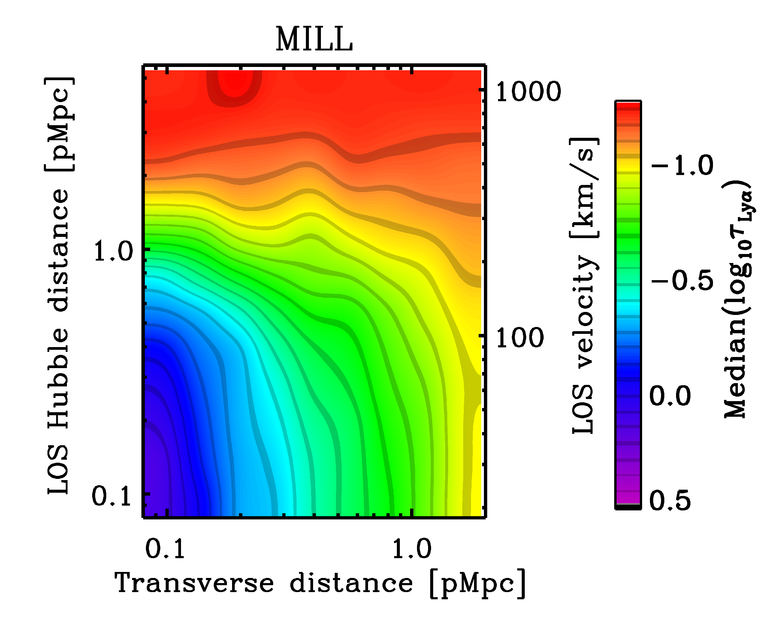}}
\caption{Median 2-D distribution of Ly$\alpha$ absorption signal in the observations of \citet[][their Figure 5]{Rakic2012}, and near haloes with $M_{\rm min}>10^{11.5}$ at $z=2.25$ using the \emph{REF}, \emph{AGN}, \emph{WML4}, and \emph{MILL} model, from left to right respectively, using 100 $h^{-1}$ cMpc simulation boxes.    \label{models}}  
 \end{figure*}
}

\def\figcb{
\begin{figure}
\begin{centering} 
\includegraphics[trim = 30mm 130mm 30mm 40mm, clip, width=0.5\textwidth]{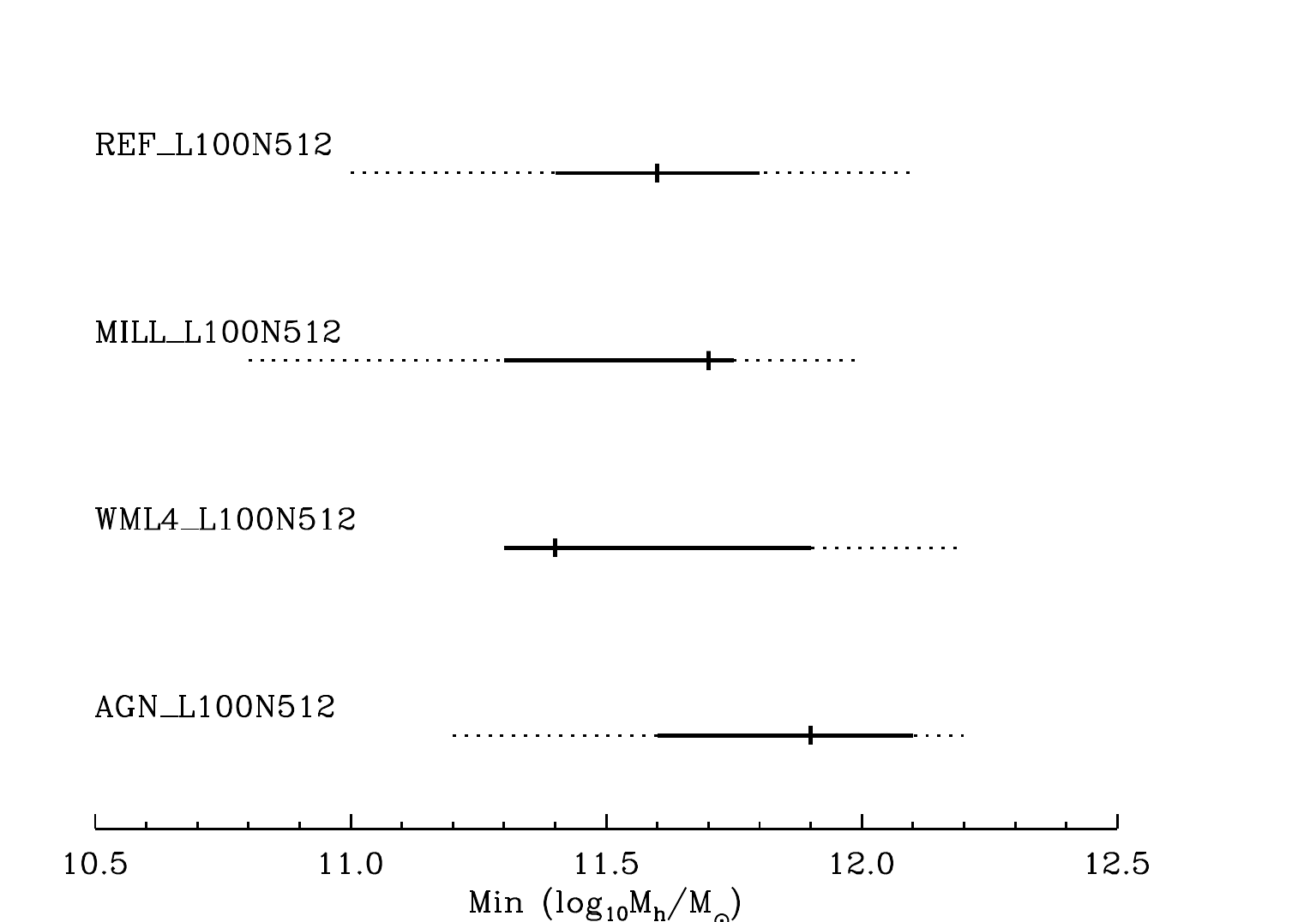}
\caption{Comparison of halo mass measurements based on comparison of the observations with the \emph{REF},  \emph{MILL}, \emph{WML4},  and \emph{AGN} models, using 100 $h^{-1}$ cMpc simulation boxes. Vertical lines show the inferred minimum halo mass, full lines the 1$\sigma$ confidence intervals, and dotted lines the 2$\sigma$ confidence intervals. All measurements agree within their 1$\sigma$ errors, which implies that the method is robust to changes in the feedback prescription and in the assumed cosmology.  \label{masscomparison}}  
 \end{centering} 
 \end{figure}
}

\def\figca{
\begin{figure}
\begin{centering} 
\includegraphics[width=0.45\textwidth]{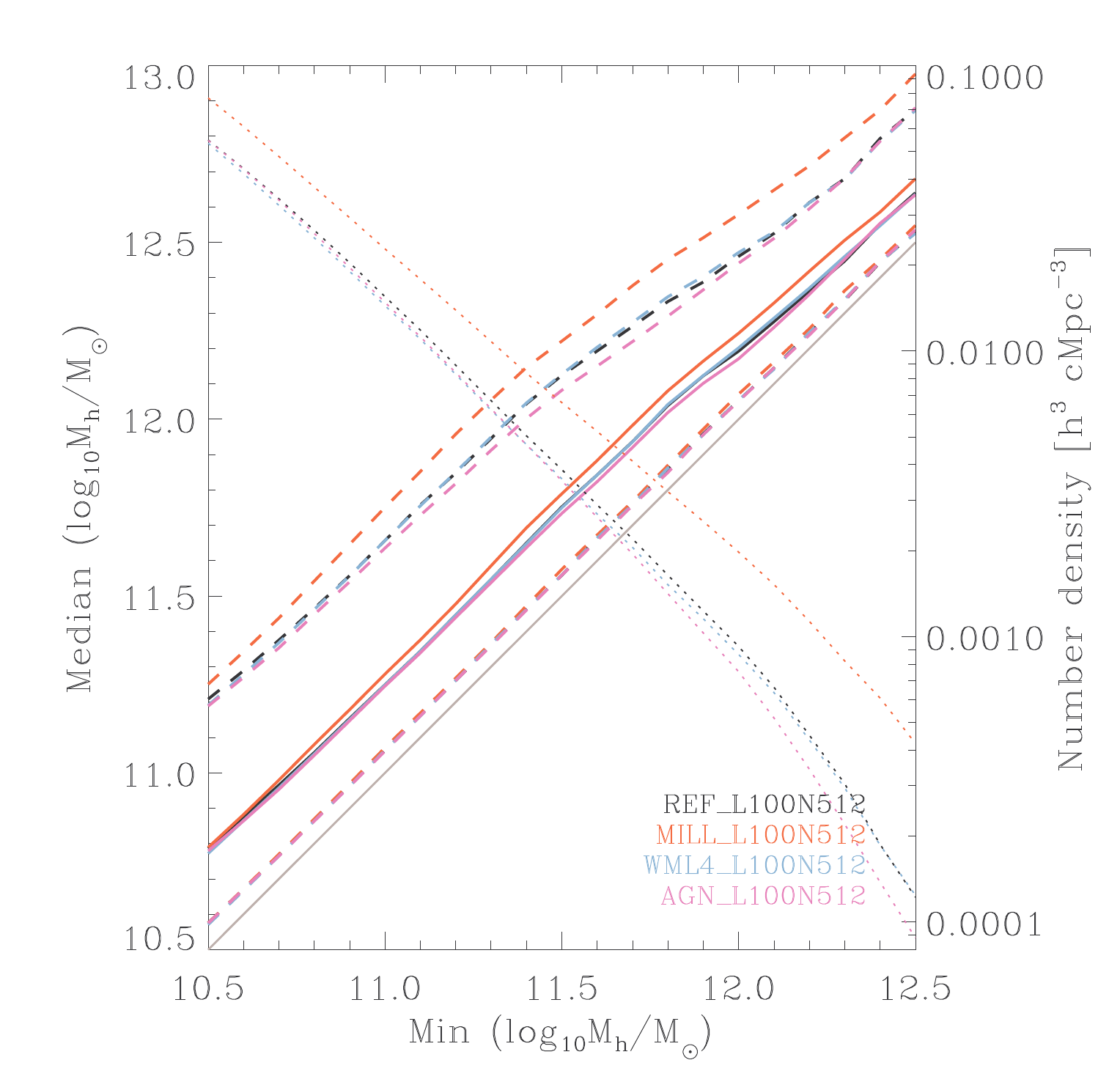}
\caption{Median halo mass above a given minimum halo mass (rising thick colored curves), for different models, as measured from 100 $h^{-1}$ cMpc boxes, with the corresponding 15.9th and 84.1th percentiles (dashed rising curves above and below the median curves). The gray line shows where median halo mass is equal to the minimum halo mass. The declining dotted curves show the corresponding comoving number densities of haloes above a given minimum mass (right y-axis).  \label{medianhalomass}}  
 \end{centering} 
 \end{figure}
}

\def\figd{
\begin{figure*}
\centering
\resizebox{0.202\textwidth}{!}{\includegraphics*[trim = 0mm 0mm 85mm 0mm, clip]{F2.png}}\resizebox{0.17\textwidth}{!}{\includegraphics*[trim = 30mm 0mm 75mm 0mm, clip]{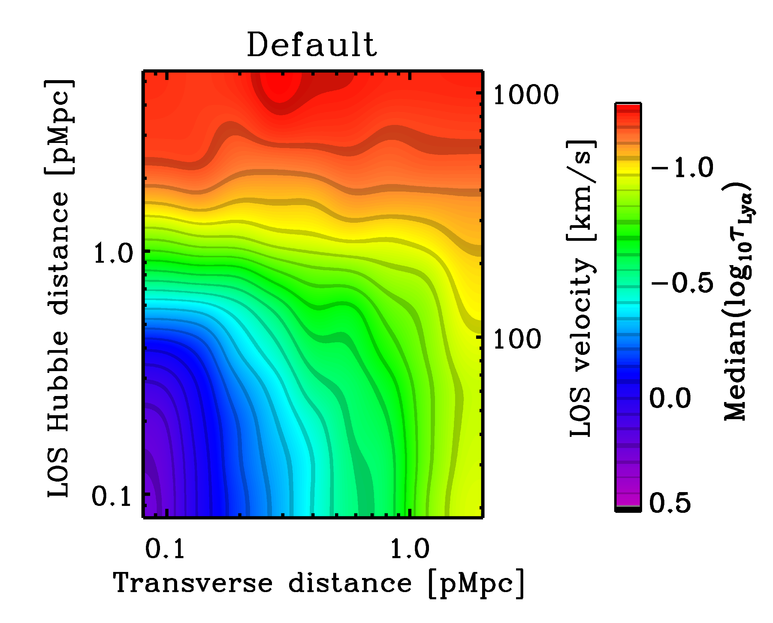}}\resizebox{0.17\textwidth}{!}{\includegraphics*[trim = 30mm 0mm 75mm 0mm, clip]{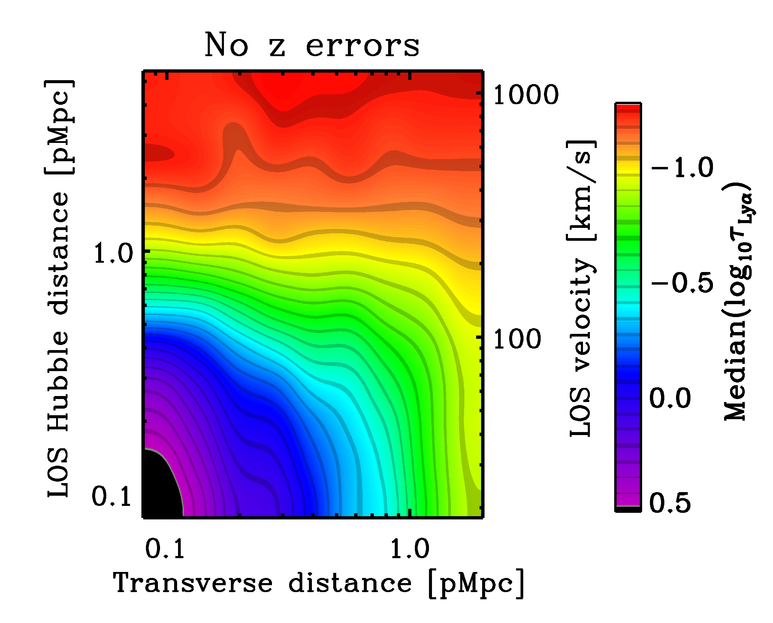}}\resizebox{0.17\textwidth}{!}{\includegraphics*[trim = 30mm 0mm 75mm 0mm, clip]{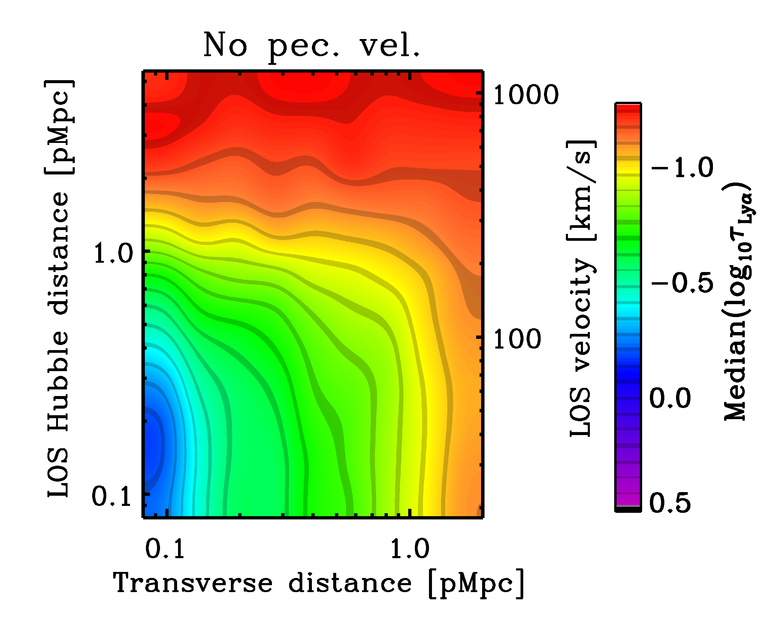}}\resizebox{0.242\textwidth}{!}{\includegraphics*[trim = 30mm 0mm 0mm 0mm, clip]{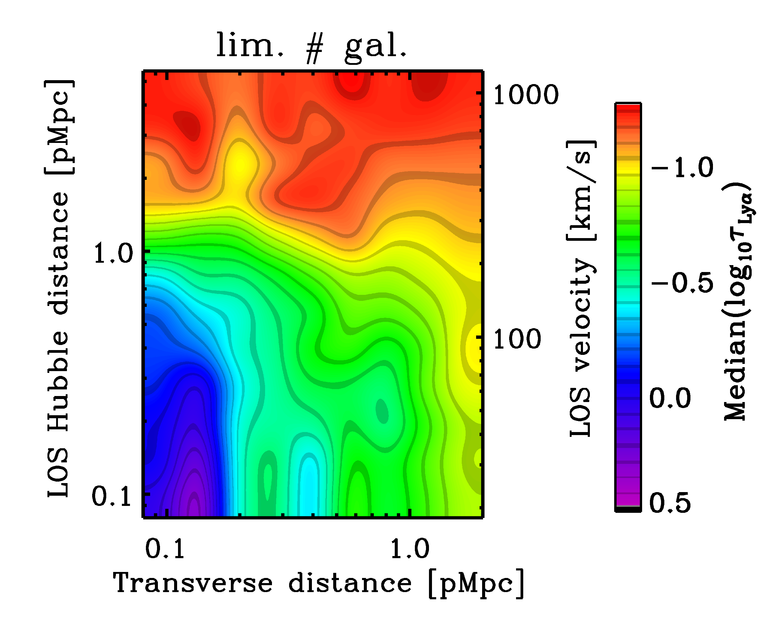}}
\caption{2-D Ly$\alpha$ absorption maps of the observations of \citet[][leftmost panel]{Rakic2012}, and for haloes with $M_{\rm h}>10^{11.5}\Msun$ at $z=2.25$ in the 50 $h^{-1}$ cMpc \emph{REF} simulation (remaining panels). From the second left to the right, the panels show: the default case (includes errors in galaxy redshifts,  and takes peculiar velocities into account), no redshift errors (the median optical depth at small impact parameters exceeds the maximum of the colour scale and appears black), ignoring peculiar velocities, and using the same number of galaxies per impact parameter bin as in the observations of R12. Redshift errors increase redshift space distortions at small impact parameters and decrease the absorption, while peculiar velocities act to compress the signal along the LOS on large scales, elongate it along the LOS for small impact parameters, and increase the absorption. \label{2D_2}}  
 \end{figure*}
}

\def\figda{
\begin{figure*}
\resizebox{0.202\textwidth}{!}{\includegraphics*[trim = 0mm 0mm 85mm 0mm, clip]{F2.png}}\resizebox{0.17\textwidth}{!}{\includegraphics*[trim = 30mm 0mm 75mm 0mm, clip]{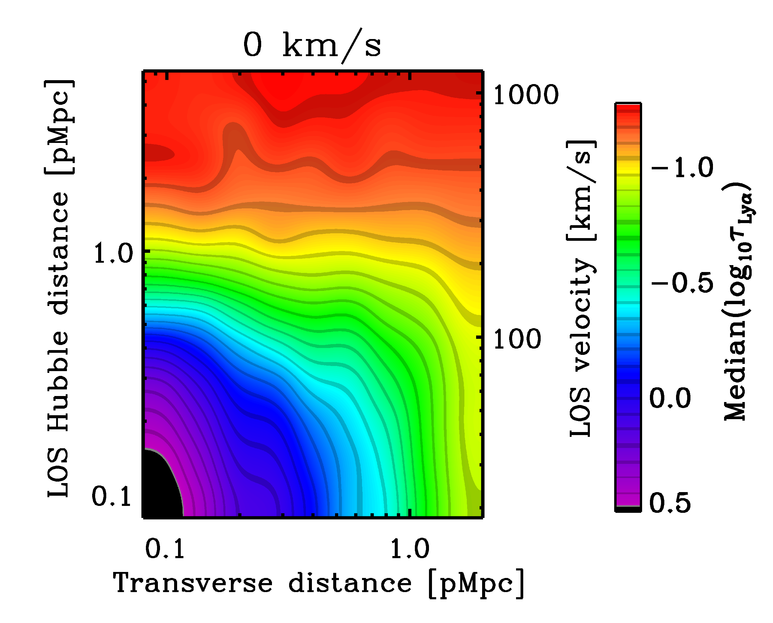}}\resizebox{0.17\textwidth}{!}{\includegraphics*[trim = 30mm 0mm 75mm 0mm, clip]{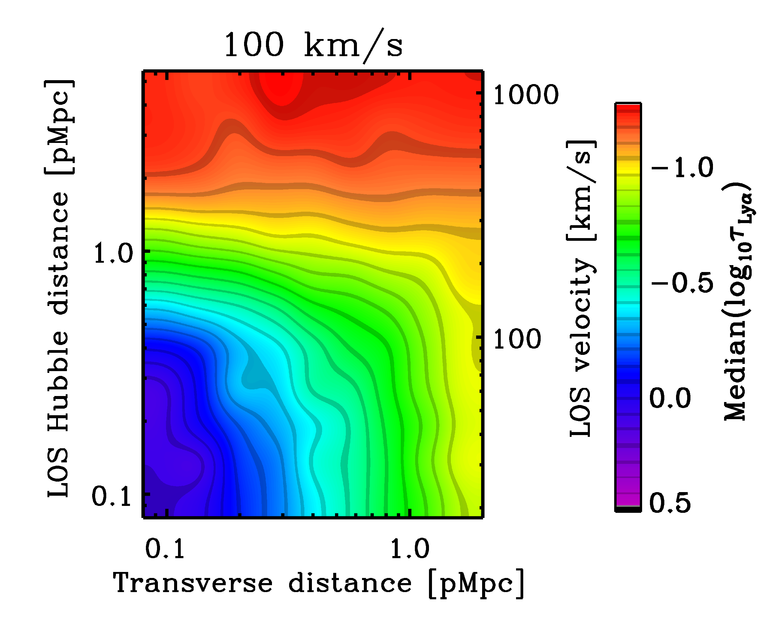}}
\resizebox{0.17\textwidth}{!}{\includegraphics*[trim = 30mm 0mm 75mm 0mm, clip]{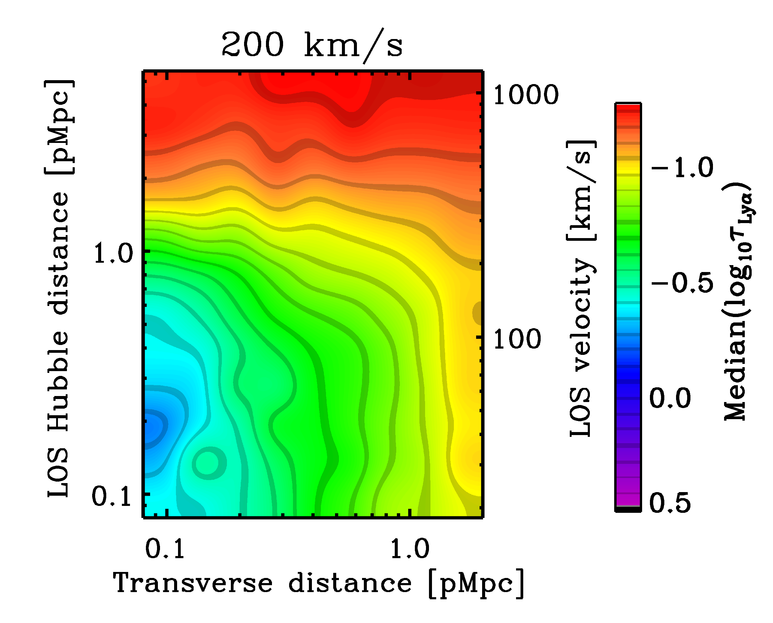}}\resizebox{0.242\textwidth}{!}{\includegraphics*[trim = 30mm 0mm 0mm 0mm, clip]{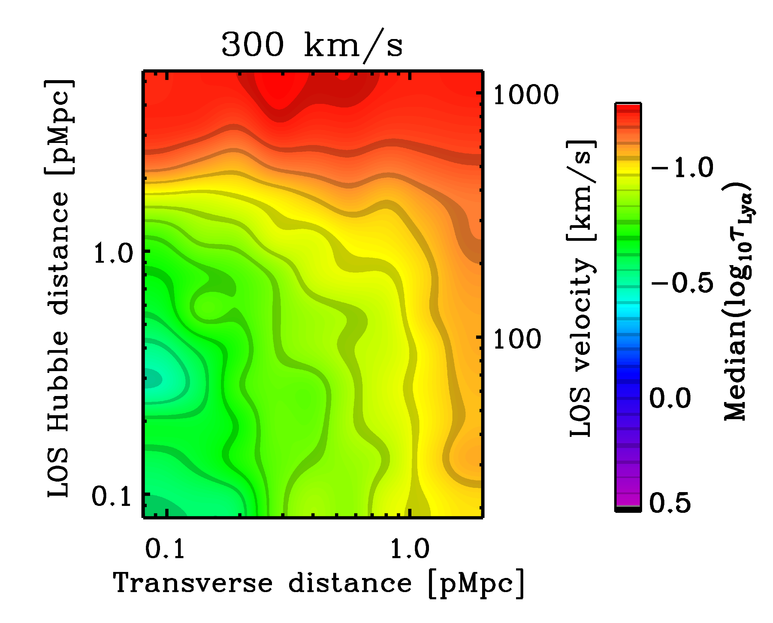}}
\caption{Similar to Figure~\ref{2D_2}, the first panel shows the observations of R12, and in the rest of panels we vary the size of the errors in the galaxy redshifts in the simulations (50 $h^{-1}$ cMpc box, \emph{REF} simulation): $\sigma=0$ (the median optical depth at small impact parameters exceeds the scale maximum, and appears black), 100, 200, and $300\rm\, km\, s^{-1}$, from left to right. Adding errors with $\sigma\gtrsim200\rm\, km\, s^{-1}$ masks the large-scale ($\gtrsim1.5$ pMpc) compression of the absorption distribution along the LOS relative to that transverse to the LOS. \label{2D_zerrors}}  
 \end{figure*}
}

\def\fige{
\begin{figure*}
\includegraphics[width=\textwidth,trim=1.cm 2.95cm 1.75cm 0cm,clip]{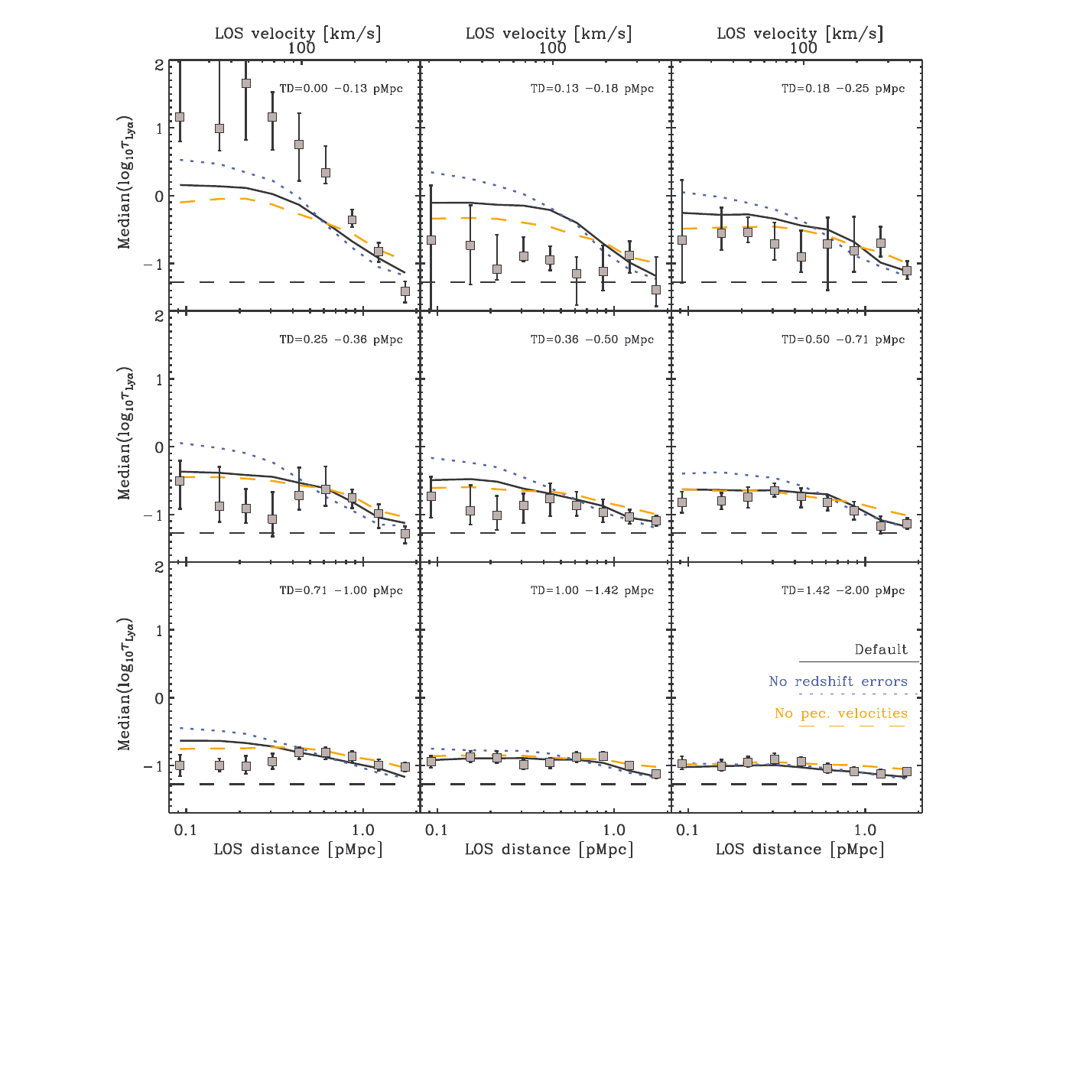}
\caption{Cuts along the LOS through the panels of Figure~\ref{2D_2} ($\rm M_{\rm h}>10^{11.5}\Msun$). The gray symbols show the observations of R12. The horizontal dashed line indicates the median optical depth of all pixels. Black solid curves show the default case (includes errors in galaxy redshifts,  and takes peculiar velocities into account), blue dotted curves show the results without redshift errors, and orange dashed curves the case where peculiar velocities are ignored. Please note that the mass measurement above is done by comparison of the observed and simulated maps for the LOS separations of 154-616 $\rm km\, s^{-1}$, while this figure extends only to 436 $\rm km\, s^{-1}$.  \label{2D_LOScuts_2}}  
 \end{figure*}
}

\def\figf{
\begin{figure*}
\includegraphics[width=\textwidth,trim=1.cm 2.95cm 1.75cm 0cm,clip]{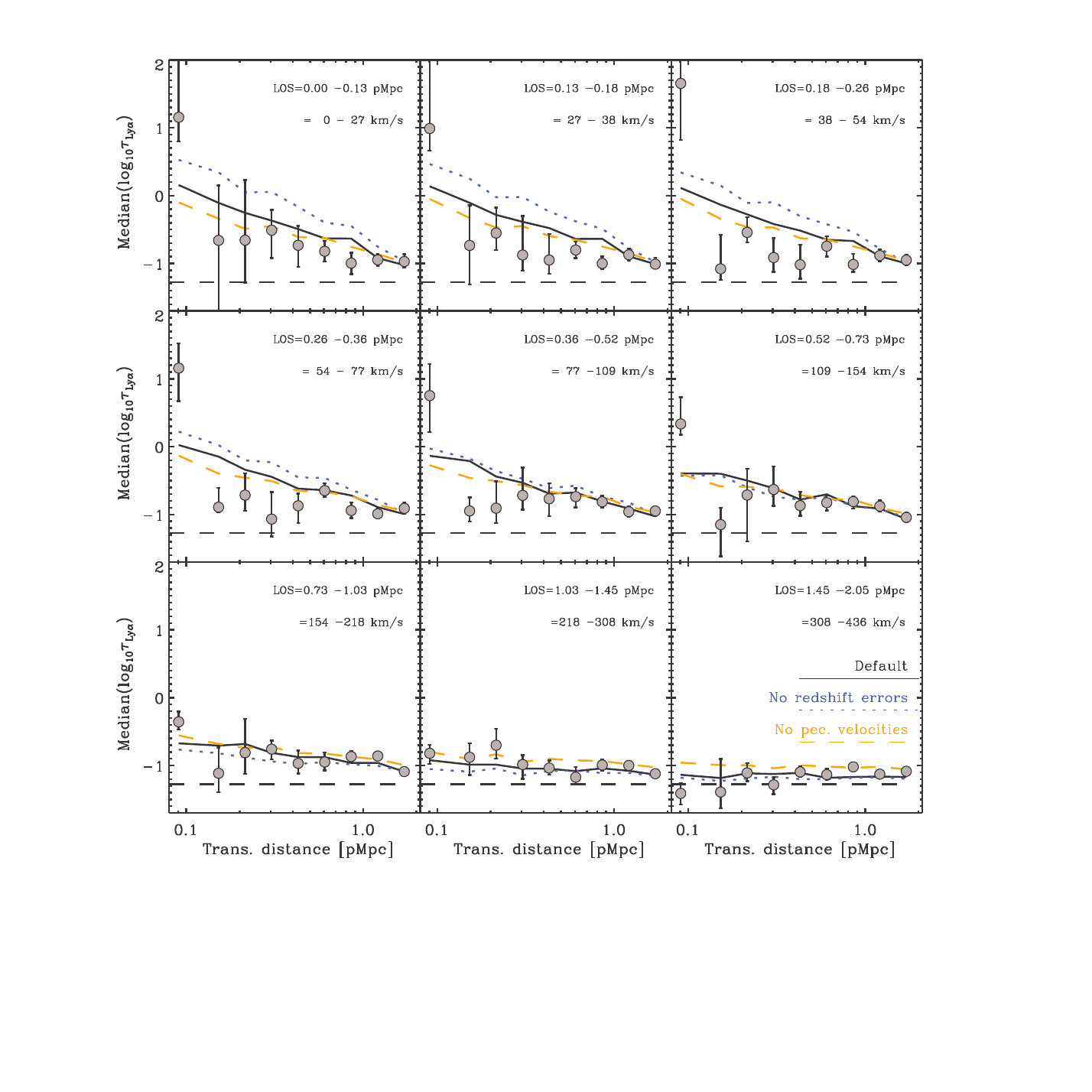}
\caption{Similar to Figure~\ref{2D_LOScuts_2}, but showing cuts in the transverse direction through the panels of Figure~\ref{2D_2} ($\rm M_{\rm h}>10^{11.5}\Msun$). \label{2D_TDcuts_2}}  
 \end{figure*}
}

\section{Introduction}
The mass of the host dark matter halo is believed to be a key factor determining galaxy evolution. Measuring halo masses is therefore an important goal of observational astronomy. For clusters of galaxies, the employed techniques include measurements of the Sunyaev-Zeldovich effect \citep[e.g.][]{Sunyaev1972,Motl2005}, cluster X-ray temperature and luminosity \citep[e.g.][]{Reiprich2002}, strong lensing \citep[e.g.][]{Gavazzi2007}, weak lensing \citep[also called galaxy-galaxy lensing; e.g.][]{Kaiser1993}, and kinematics of galaxies \citep[e.g.][]{Zwicky1937}. Measuring dark matter halo masses of individual galaxies is more difficult and usually requires statistical methods such as stacking combined with weak lensing \citep[e.g.][]{Mandelbaum2006}, satellite kinematics \citep[e.g.][]{More2011}, clustering analysis  \citep[e.g.][]{Kaiser1984}, or abundance matching \citep[e.g.][]{Conroy2006}. Note, however, that \citet{Haas2011}  found that the most popular environmental indicators (e.g.\ the number of galaxies within a given distance, or the distance to the Nth nearest neighbour)  correlate strongly  with halo mass and can thus be used to estimate halo masses of single galaxies.

While weak lensing  has been successful in estimating dark matter halo masses for  low-redshift galaxies, it is not a viable method for high-redshift objects since the lensing cross-section peaks half way towards the background galaxy, and is thus very small at $z\geq2$. Furthermore, the number of both foreground and background galaxies becomes too low to obtain statistical samples, and measuring shapes of galaxies becomes increasingly hard. 

As known from structure formation studies, more massive dark matter haloes cluster more strongly than lower mass haloes, and on scales greater than the haloes, the galaxies inhabiting such haloes cluster similarly as their host haloes. Matching the observed clustering properties of a given galaxy population with the clustering of haloes in N-body simulations leads to estimates of the most likely host halo masses. This method has facilitated some important insights  into the properties of galaxies, and their likely descendants \citep[e.g.][]{Adelberger1998,Adelberger2005b,Quadri2007,Conroy2008}.

Here we measure galaxy halo masses from \ion{H}{I} Ly$\alpha$ absorption profiles. Numerous studies have shown that the mean absorption of the light from a background object (e.g.\ a bright QSO) is increased when the sightline passes near a foreground galaxy, both observationally \citep[e.g.][]{Chen1998, Chen2001, Penton2002, Adelberger2003,Adelberger2005,Crighton2010,Prochaska2011, Rakic2011,Rakic2012,Rudie2012}, and using simulations \citep[e.g.][]{Croft2002,Kollmeier2003,Bruscoli2003,Desjacques2004,Kollmeier2006,Desjacques2006,Tasker2006,Bertone2006,Stinson2012,Ford2012}. At high-redshift this enhancement can be detected statistically out to distances of several proper Mpc (pMpc). \citet{Kim2008} proposed to use such absorption profiles to measure halo masses of foreground QSOs and galaxies.  Somewhat related to this method is the work of  \citet{FG2008}, who used simulations to show that the Ly$\alpha$ optical depth statistics in the QSO proximity zones depend on the dark matter halo masses of the QSO hosts. 

We apply a modified version of this method to the observations from \citet[][hereafter R12]{Rakic2012}. R12 measured the \ion{H}{I} Ly$\alpha$ absorption distribution around UV selected star-forming galaxies from the Keck Baryonic Structure Survey, KBSS (Steidel et al., in preparation), at $\langle z\rangle\approx2.36$. Through the analysis of the absorption spectra of background QSOs they found enhanced Ly$\alpha$ absorption within $\approx 3$ proper Mpc (pMpc) from galaxies. They also presented the 2-D (i.e.\  along the line of sight, LOS, and in the transverse direction) absorption distribution  and they measured a slight compression of the signal along the LOS on scales $\gtrsim1.5\,\rm pMpc$. This compression suggests large-scale infall of gas into the potential wells occupied by the galaxies.

This  observational study is particularly suitable for testing the method because the dark matter halo masses of the galaxies are already known from clustering analyses \citep[e.g][]{Adelberger2005b,Conroy2008,Trainor2012}. \citet{Conroy2008} found that every halo at $z\sim2$ with $M_{\rm h}>10^{11.4}h^{-1}\Msun$ hosts a star-forming galaxy from the photometric sample from which the galaxies used in R12 were drawn. This result was obtained using N-body simulations with the WMAP year-1 cosmology. Using the WMAP year-3 cosmology (which is what the simulations in this paper are based on), they found an 0.3-0.4 dex smaller minimum halo mass. Using a spectroscopic sample that is almost identical to the one used by R12, \citet{Trainor2012} found that these galaxies reside in halos with $M_{\rm h}>10^{11.7}\Msun$, using WMAP year-7 cosmological parameters.

The paper is organized as follows. We describe the simulations in Section~\ref{SimulationsMass}. The method for measuring halo mass is presented in Section~\ref{MethodMass}. We discuss the origin of the observed redshift space distortions seen in the 2-D absorption maps from R12 in Section~\ref{zdistortions}. Finally, we conclude in Section~\ref{Summary}. The appendix demonstrates that our results are converged with respect to the numerical resolution and the box size of our simulations.

Throughout the text proper distances are denoted as pMpc, and comoving as cMpc. 

\section{Simulations}\label{SimulationsMass}
We use a subset of simulations from the suite of OverWhelmingly Large Simulations \citep[\textsc{OWLS}; ][]{Schaye2010}. The simulations were run using a modified version of \textsc{gadget-3} \normalfont \citep[last described in][]{Springel2005}, which is a Lagrangian (SPH) code.  These simulations include star formation \citep{Schaye2008}, supernova (SN) feedback \citep{DallaVecchia2008}, radiative heating and cooling via 11 elements (hydrogen, helium, carbon, nitrogen, oxygen, neon, magnesium, silicon, sulphur, calcium, and iron) in the presence of the cosmic microwave background and the \citet{HaardtMadau2001} model for the ionizing background radiation from galaxies and quasars \citep{Wiersma2009}, and chemodynamics \citep[abundances of elements released by type Ia and type II SNe, stellar winds, and asymptotic giant branch stars; ][]{Wiersma2009b}. The assumed initial mass function (IMF) is that of \citet{Chabrier2003}, with stars ranging in mass from 0.1 to $100\,M_{\odot}$. Stars with masses greater than $6\Msun$ die as SNII, injecting $\sim10^{51}\rm erg$ of energy into their surroundings. About 40\% of this energy is injected kinetically, where particles are kicked at a velocity of $600~\rm km\, s^{-1}$. The mass loading parameter, $\eta$, is set to 2, i.e.\  the total mass of the particles that are kicked is on average equal to twice the mass of the star particle. The box sizes and (the initial) baryonic and dark matter particle masses of the simulations used here  are specified in Table~\ref{simulations.tab}. The cosmological parameters used in the simulations are from the Wilkinson Microwave Anisotropy Probe (WMAP) year-3 results \citep{Spergel2007}, $\{\Omega_{\rm m}, \Omega_{\rm b}, \Omega_{\Lambda}, \sigma_{8} , n_{\rm s} , h\}=\{0.238, 0.0418, 0.762, 0.74, 0.951, 0.73\}$. 

We use the `reference' (\emph{REF}) model, the model with AGN feedback (\emph{AGN}), the model using the `Millennium cosmology' and more efficient feedback (\emph{MILL}), and a model with the same SN feedback prescription as \emph{MILL} but the WMAP year-3 cosmology (\emph{WML4}). All of the models  are described in detail by \citet{Schaye2010}; here we provide only short descriptions and a brief summary can be found in Table~\ref{unscaledtau.tab}. 

The \emph{REF} and the \emph{AGN} models are identical, except that the latter includes prescriptions for the growth of supermassive black holes and for AGN feedback.  These processes are implemented using modified versions of those of \citet{Springel2005}, and are described in \citet{Booth2009}. Black holes inject 1.5 \% of the rest-mass energy of accreted gas in the form of heat into the surrounding matter. \citet{Booth2009,Booth2010} show that this efficiency results in excellent agreement with the observed, $z=0$ scaling relations between black hole masses and the properties of their host galaxies.  We chose this simulation to test the effect of extreme feedback on our results as  it could potentially influence the physical properties (e.g.\ the density and temperature) of the absorbing gas near galaxies.

To test the effect of the cosmological parameters, and to facilitate comparisons with previous studies that used older estimates of these parameters, we employ the `Millennium cosmology' \citep{Springel2005b} in our \emph{MILL} simulation. The \emph{MILL} simulation  uses the cosmological parameter values $\{\Omega_{\rm m}, \Omega_{\rm b},$ $\Omega_{\Lambda}, \sigma_{8}, n_{\rm s}, h\}=$ $ \{0.25, 0.045, 0.75, 0.9, 1.0, 0.73\}$. As found by e.g.\ \citet{Conroy2008} and \citet{Trainor2012}, the lower values of $\sigma_{8}$ that are implied by more recent WMAP results cause the lower mass dark matter haloes to cluster more strongly than in models with higher $\sigma_{8}$. Hence the minimum mass of haloes implied by the observed clustering of a galaxy population is lower. The faster growth of structure in the \emph{MILL} simulation also causes higher predicted global star-formation rate (SFR) densities. The mass loading factor for the SN driven winds is therefore doubled in comparison with the \emph{REF} model, to $\eta=4$. This is done in order to match the peak in the observed star-formation history (SFH). We note that the WMAP year-7, estimates of cosmological parameters \citep{Komatsu2011}, $\{\Omega_{\rm m}, \Omega_{\rm b}, \Omega_{\Lambda}, \sigma_{8} , n_{\rm s} , h\}=\{0.272, 0.0455, 0.728, 0.81, 0.967, 0.704\}$, prefer a value of $\sigma_{8}$  that is intermediate between those from WMAP year 1 and WMAP year 3, while the value for $\Omega_{\rm m}$ is closer to that from WMAP1.

\emph{WML4} is identical to the \emph{REF} model, except that the mass loading factor is the same as in the \emph{MILL} simulation, which allows us to separate the effects of cosmological parameters and feedback prescriptions. 

\begin{table*}
   	\begin{minipage}{.78\linewidth}
      	\caption{Numerical parameters. The columns give the comoving size
  	of the box $L$, the total number of particles per component $N$ (dark
  	matter and baryons), the initial baryon particle mass $m_{\rm b}$, the dark matter
  	particle mass $m_{\rm dm}$, and the maximum proper gravitational softening length (above $z=3$ it is fixed in comoving coordinates).}
     	\begin{tabular}{cccccc}

        	\hline
        	\noalign{\smallskip}
	Simulation & $L$ 		& $N$ & $m_{\rm b}$   & $m_{\rm dm}$ & Max. soft. length \\
			  &  [$h^{-1}$cMpc] &   	   &  [$\hMsun$] &  [$\hMsun$] & [$h^{-1}\rm ckpc$]\\
	
	\noalign{\smallskip}
	\hline
	\noalign{\smallskip}
	L050	N512 &  50 & $512^3$ & $1.1 \times 10^7$ & $ 5.1 \times 10^7$ & 1 \\
	L100	N512 & 100 & $512^3$ & $8.7 \times 10^7$ & $ 4.1 \times 10^8$  & 2\\
        	\noalign{\smallskip}
	\hline
      	\end{tabular}
      	\label{simulations.tab}
    	\end{minipage}
\end{table*}

\subsection{Extracting sightlines from the simulations}\label{SimulatingObservations} 
We use the $z=2.25$ snapshot, as this is close to the peak in the redshift distribution of the galaxy sample used in R12 ($z\approx 2.36$), and the snapshots of the simulations are saved only every $\Delta z=0.25$. Although the difference between $z=2.25$ and $z=2.36$ is comparable to a change of 3\% in $\sigma_8$, in Section~\ref{Models} we show that using simulations with different cosmological parameters, with values of $\sigma_8$ differing by $22\%$, does not lead to significantly different results.

The masses and locations of the dark matter haloes in the simulations are determined using a spherical overdensity criterion as implemented in the  \textsc{SubFind} algorithm  \citep{Springel2001}. \textsc{SubFind} finds the radius, $r_{200}$, and mass, $M_{\rm h}$, of a spherical halo centered on the  potential minimum of each identified halo, so that it contains a mean density of 200 times the critical density of the Universe at a given redshift.

We begin by defining a number of halo samples, each characterized by one parameter, the minimum halo mass, $M_{\rm min}$. Each sample consists of all haloes with $M_{\rm h}>M_{\rm min}$. We vary $M_{\rm min}$ in steps of 0.1 dex between  $M_{\rm h}=10^{10.5}-10^{12.5}\Msun$ and thus consider a total of 20 different halo samples. For each of these samples, we fire 12,500 sightlines through the simulation volume such that each one passes within 5~pMpc of a randomly chosen halo  (which ensures that we sample the halo mass function correctly). There are 500 sightlines for every 200 pkpc impact parameter bin. Given that there are, e.g.\  $\sim20,000$ haloes with mass above $10^{11}\Msun$, or $\sim700$ haloes with mass higher than $10^{12}\Msun$ in the 100 $h^{-1}$ cMpc box (see Figure~\ref{medianhalomass} for the number densities of haloes above a given mass), a single halo features only once in most distance bins. Although sightlines are always parallel to a side of the box, the axis is chosen randomly. In addition, the position angle with respect to galaxies is also random.
\figca

By setting only the minimum halo mass, we implicitly assume that each halo above a given mass limit hosts a galaxy from the observed population. Objects that are potentially missed by the selection based on rest-frame UV colors, e.g.\  massive red galaxies identified through their rest-frame optical colors \citep[e.g.][]{Franx2003}, have space densities that are  more than an order of magnitude smaller than the rest-frame UV selected galaxies used here, and are thus expected to have little impact on our statistics. We verified that using overlapping bins in halo mass with  minimum and maximum masses spaced by 1 dex yields very similar results to setting only the lower mass limit.  This is due to the shape of the halo mass function, where the typical (median) mass above a given mass limit is only 0.2-0.3 dex higher than the imposed minimum (Figure~\ref{medianhalomass}).

The synthetic spectra were created using the package SPECWIZARD written by Schaye, Booth, \& Theuns using the procedure described in  Appendix A4 of \citet{Theuns1998}. The QSO spectra from R12 were taken with Keck I/HIRES \citep{Vogt1994}, have a spectral resolution of $\rm FWHM\approx8.5\,\rm km\, s^{-1}$, and were rebinned to $2.8\rm\, km\, s^{-1}$ pixels. We convolved our simulated spectra with a Gaussian with a $\rm FWHM=6.6\rm\, km\, s^{-1}$, and rebinned to $1.4\rm\, km\, s^{-1}$ pixels. We verified that the results are almost identical if we use the exact match to observed spectra, i.e.\ $\rm FWHM\approx8.5\,\rm km\, s^{-1}$, rebinned to $2.8\rm\, km\, s^{-1}$, instead of $\rm FWHM=6.6\rm\, km\, s^{-1}$, and rebinned to $1.4\rm\, km\, s^{-1}$. We added Gaussian noise with a signal to noise (S/N) ratio of 100 to the spectra, which matches the typical S/N of the observations. 

Due to uncertainties in the intensity of the ionizing background radiation, it is common practice to scale the optical depth in simulated spectra  to match the mean/median absorption level of observed spectra at a given redshift.  In order to match the median optical depth  in the observed spectra of  \citet[][median log$_{10}\tau_{\rm Ly\alpha}=-1.27_{-0.03}^{+0.02}$]{Rakic2012},  we select 1000 random sightlines and find the optical depth of all pixels from spectra along those sightlines, from which we then determine the relevant scaling factor. This is done separately for each simulation box used in this study and the same factor is then used for all the spectra drawn from the corresponding box. The unscaled median optical depths for the different simulations are listed in Table~\ref{unscaledtau.tab}. The mean normalized flux in the sample of observed QSO spectra is 0.806. The formula for the effective Ly$\alpha$ optical depth from \citet{Schaye2003} suggests that the mean flux at $z=2.36$ is 0.802($\pm0.008$), which is consistent with the observations, so we conclude that the IGM probed by the observed QSO sightlines from R12 is representative for the considered redshift. 

\begin{table*}
  \begin{minipage}{.78\linewidth}
  \caption{List of main physics variations employed in the used models and an unscaled median optical depth for each simulation. }
  \begin{tabular}{l|rlr}

  \hline
  \noalign{\smallskip}
  Models & Box size [$h^{-1}$cMpc] & Description & log$_{10}\tau_{\rm Ly\alpha}$  \\ 
	\noalign{\smallskip}
	\hline
	\emph{REF}  & 100	&	Reference model &-1.34 \\
	\emph{REF}  & 50 	&	Reference model &-1.40\\
	\emph{AGN}	& 100 &    Includes AGN &-1.37\\	 
	\emph{MILL} 	& 100 &	Millennium simulation cosmology, $\eta = 4$ (twice the SN energy of \emph{REF}) &-1.48\\
	\emph{WML4}	&  100 &   Wind mass loading $\eta = 4$ (twice the SN energy of \emph{REF})&-1.35\\
   \noalign{\smallskip}
   \hline
   \end{tabular}
   \label{unscaledtau.tab}
   \end{minipage}
\end{table*}

\subsection{Resolution tests}
The \emph{REF} model was run in 25, 50, and 100 $h^{-1}$cMpc boxes, while the rest of the simulations are only available in 25 and 100 $h^{-1}$cMpc boxes. The smallest box size does not contain sufficient high-mass haloes and is therefore not suitable for our study. The 50 $h^{-1}$cMpc simulation box samples the relevant range of the mass function well, and  we used this box size to test convergence  by comparing it with the 100 $h^{-1}$cMpc box. The latter  samples the high-end of the mass function well, but at the low-mass end it resolves haloes with $M_{\rm h}<10^{10.5}\Msun$ with less than $10^2$ dark matter particles (for the simulation with $512^3$ particles), which is insufficient to robustly identify dark matter haloes \citep[e.g.][]{Jenkins2001}. However, comparison of the \emph{REF} model with observations suggests that the relevant mass range is $M_{\rm h}>10^{11}\Msun$, which means that 100 $h^{-1}$cMpc simulation boxes are appropriate for our particular problem. We show  convergence tests in the Appendix.

\section{Measuring halo masses}\label{MethodMass}
In this section we show the results of matching 2-D \ion{H}{I} Ly$\alpha$ absorption maps, and 1-D cuts through such maps, to those from the simulations for different minimum halo masses. R12 show 2-D maps of the median \ion{H}{I} Ly$\alpha$ optical depth around galaxies, i.e.\ the optical depth distribution as a function of transverse and LOS separation from galaxies, where distances along the LOS are calculated from velocity separations under the assumption that they are due to the Hubble flow and that there are no peculiar velocities. Given that maps are built by combining many galaxies, in the absence of peculiar velocities the absorption distribution would appear isotropic (assuming the Universe is isotropic in a statistical sense). Therefore, any departure from isotropy in 2-D absorption maps can be attributed to peculiar velocities and/or errors in galaxy redshifts.

R12 used the pixel optical depth method \citep[e.g.][]{CowieSongaila1998,Ellison2000,Schaye2000,Aguirre2002} to recover the \ion{H}{I} Ly$\alpha$ optical depth in each pixel of a QSO spectrum. Each galaxy in the field of a QSO is then associated with an array of pixels with varying velocity separation, but at a fixed impact parameter. This is done for each QSO field, and then all the QSO fields are combined to build a galaxy-centered 2-D map, where both transverse and LOS distances are binned logarithmically, and where only absolute LOS velocity difference is taken into account. The simulated maps are built in an analogous way, where instead of galaxies we use dark matter haloes. 

For reference, the number of galaxies in the observations per transverse bin depends slightly on the velocity difference (see R12 for details). From small to large impact parameters, the number of galaxies contributing to the 1st (9th) velocity bin is 14 (16), 8 (8), 11 (12), 22 (23), 47 (48), 58 (62), 96 (99), 175 (180), 202 (211).

\subsection{Measuring halo mass from 2-D absorption maps}

The first panel of Figure~\ref{2D_masses} shows the observed 2-D \ion{H}{I} Ly$\alpha$  absorption map from R12. The remaining panels show, from left to right, maps from simulations for minimum halo masses of  $M_{\rm min}=10^{10.5}$, $10^{11.5}$, and $10^{12.5}\Msun$. The distance bins in these maps are logarithmically spaced by 0.15 dex, both in the transverse direction and along the LOS. We added  errors to the simulated galaxy redshifts, to mimic those from the observations \citep{Steidel2010}: to a random 10$\%$ of the simulated redshifts we added errors drawn from a Gaussian distribution with $\sigma=60\rm\, km\, s^{-1}$ (mimicking errors in redshifts measured from nebular emission lines), and to the remaining 90$\%$ of redshifts errors with $\sigma=125\rm\, km\, s^{-1}$ (imitating errors measured from rest-frame UV absorption and emission lines). 

Haloes with $M_{\rm h}>10^{10.5}\Msun$ produce too little absorption to account for the observed absorption at large impact parameters, while the more massive haloes, $M_{\rm h}>10^{12.5}\Msun$, produce too much absorption in those regions. The third panel, showing absorption near haloes with  mass $M_{\rm h}>10^{11.5}\Msun$, resembles the observed distribution the most. The observed map is noisier, however, which is not surprising given that hundreds of sightlines contribute to each bin of the simulated maps, compared to $\approx10$ in the innermost impact parameter bins of the observed map. None of the simulated maps shows enough absorption very close to galaxies (impact parameters  $b\lesssim100$ pkpc). \citet{Steidel2010} find that the absorption strength keeps rising closer to galaxies using their galaxy-galaxy pairs observations, which sample the $b<100$ pkpc region well. This suggests that there may be a true deficiency of high column gas close to galaxies in the simulations and that the discrepancy may not only be a result of small number statistics \citep[there are only 15 galaxies in the first, and 8 in the second impact parameter bin of][]{Rakic2012}. We will come back to this question in more detail below.

\figb

Figure~\ref{2D_masses} indicates the approximate mass range for which host haloes produce absorption comparable to the observations. More quantitatively, we estimate the minimum halo mass by minimizing the reduced $\chi^2$ between the observed 2-D Ly$\alpha$ absorption maps and maps from the simulations, taking into account errors on observed data points. We use only regions of maps with LOS separations smaller than $616\rm\,km\, s^{-1}$ because the absorption signal  is consistent with noise beyond this point (see Figure~\ref{2D_masses}). Different distance bins in 2-D maps are uncorrelated in the transverse direction, but they are correlated on scales of $200\rm\, km\, s^{-1}$ along the LOS (R12). We therefore use 1000 bootstrap realizations of the observed galaxy sample to estimate the errors on the halo mass, instead of using a $\Delta\chi^2$ criterion. For each bootstrap realization of the data, we find the implied minimum halo mass by comparing the resulting maps to the simulated maps, and then find the 1 and $2\sigma$ confidence intervals from the sample of bootstrap realizations. Given a grid of minimum halo masses spaced by 0.1 dex, the errors on halo masses take discrete values. we set the minimum error to half the grid spacing, i.e.\  0.05 dex, since we cannot determine errors to better than this value.

For the 50  $h^{-1}$cMpc \emph{REF} simulation the implied mass is log$_{10}M_{\rm min}/\Msun={11.2}^{+0.5+0.5}_{-0.4-0.4}$  (the first and second pairs of errors indicate $1\sigma$ and $2\sigma$ confidence intervals, respectively).  For the 100  $h^{-1}$cMpc box it is log$_{10}M_{\rm min}/\Msun={11.2}^{+0.3+0.5}_{-0.3-0.4}$. The two simulations give almost identical results, which justifies using larger boxes for the rest of the models (see also the Appendix). This also suggests that our mass measurements are not affected by cosmic variance due to the use of a single simulation realization.

Because  uncertain baryonic physics and redshift errors could affect the absorption signal close to galaxies, we also restrict the comparison to those parts of the maps where these uncertainties are expected to have a smaller impact. We test two cases where we exclude  regions with $v_{\rm LOS}<77\,\rm km\, s^{-1}$ (i.e.\ the first 4 velocity bins) and $154\,\rm km\, s^{-1}$ (i.e.\ the first 6 velocity bins), or $v_{\rm LOS}<79$ and $158\,\rm km\, s^{-1}$ for the \emph{MILL} model (due to the different cosmological parameters, the same proper distance corresponds to a different velocity separation), where $v_{\rm LOS}$ is the line of sight  velocity separation between absorbers and galaxies. The minimum halo mass  resulting from such conservative comparisons is in all cases consistent with the one inferred by comparing the complete maps. Given that the simulations seem to underproduce the level of absorption very close to galaxies in comparison with the observations (e.g.\ Figure~\ref{2D_masses}), and given the uncertain baryonic physics in the vicinity of galaxies, we choose  to compare absorption distributions from observations and simulations only at LOS velocity separations $v_{\rm LOS}>154\rm\, km\, s^{-1}$ (which is comparable to the redshift errors in the galaxy sample) by default. This implies a minimum halo mass of log$_{10}M_{\rm min}/\Msun={11.7}^{+0.2+0.5}_{-0.05-0.2}$ using the 50  $h^{-1}$cMpc box, and log$_{10}M_{\rm min}/\Msun={11.6}^{+0.20+0.50}_{-0.20-0.60}$ using the 100  $h^{-1}$cMpc box. 

Figure~\ref{2D_difference} shows the absolute difference between the observed map and the map for the 100  $h^{-1}$ cMpc \emph{REF} simulation, in units of the observational error. The shaded regions were not used for the measurement of halo mass. The difference between the observed and simulated absorption is $\lesssim1\sigma$ in the majority of the distance bins in the regions used for the comparison.

As a test of robustness, we check the effect of removing the innermost transverse distance bins, i.e. galaxies at small impact parameters. As already mentioned, the innermost transverse distance bins contain fewer galaxies,  which could mean that the error bars on the observed absorption profile in those bins are not as robust as those for the larger impact parameters. Considering regions at impact parameters $>0.13$ pMpc (i.e.\ excluding the innermost transverse distance bin) and LOS velocity separations $154<v_{\rm LOS}<616 \rm\, km\, s^{-1}$, implies a minimum halo mass of log$_{10}M_{\rm min}/\Msun={11.8}^{+0.1+0.4}_{-0.4-0.8}$. The minimum halo mass for  impact parameters $>0.18$ pMpc  (i.e.\ excluding the first two transverse distance bins) and LOS velocity separations $154<v_{\rm LOS}<616 \rm\, km\, s^{-1}$ is log$_{10}M_{\rm min}/\Msun={11.8}^{+0.3+0.4}_{-0.4-0.8}$. Both results are consistent with the default case within 1$\sigma$.

Finally, we check the sensitivity of the results to the scaling of optical depth in simulated spectra to match the median optical depth of observed spectra from R12. The 1$\sigma$ confidence interval of the median optical depth in the observed spectra is log10(median ${\tau_{Ly\alpha}}$)=$-1.27_{-0.03}^{+0.02}$, as calculated by bootstrapping the QSO spectra. We scale the simulated spectra to the lower and upper limits on the median optical depth, and then remeasure the  halo mass. Scaling to the lower 1$\sigma$ confidence limit implies a minimum halo mass of log$_{10}M_{\rm min}/\Msun={11.8}^{+0.4+0.5}_{-0.2-0.3}$, while scaling to the upper 1$\sigma$ confidence limit yields log$_{10}M_{\rm min}/\Msun={11.6}^{+0.2+0.2}_{-0.3-0.8}$. Both are consistent with the default result at the 1$\sigma$ level.

\figba

\subsubsection{The AGN, MILL and WML4 models}\label{Models}

We use the \emph{AGN} simulation to test the effect of feedback prescriptions on the measurement. The minimum mass implied by a comparison of the observations with this simulation is  log$_{10}M_{\rm min}/\Msun={11.9}^{+0.20+0.30}_{-0.30-0.70}$. This is somewhat higher than the mass inferred from the \emph{REF} model (log$_{10}M_{\rm min}/\Msun={11.6}^{+0.20+0.50}_{-0.20-0.60}$), but still consistent within 1$\sigma$.  \citet{McCarthy2011} used this simulation to show that AGN eject large amounts of gas from the haloes at $z\sim2-3$, which may explain why we need to move to higher mass haloes to get high enough columns of neutral gas to account for the observed absorption. Nevertheless, given that the results differ by only 0.3 dex and by less than 1$\sigma$, we conclude that our method for measuring halo mass is sufficiently robust to changes in the baryonic physics prescriptions.

To quantify the impact of cosmology on our results, we compare the \emph{MILL} and \emph{WML4} simulations, which use the same subgrid physics and differ mostly in terms of $\sigma_8$, which is  0.9 for the former and 0.74 for the latter model. We find minimum halo masses of log$_{10}M_{\rm min}/\Msun={11.7}^{+0.05+0.30}_{-0.40-0.90}$ and ${11.4}^{+0.50+0.80}_{-0.10-0.10}$ for \emph{MILL} and \emph{WML4}, respectively. These estimates are consistent with each other and with the value implied by the \emph{REF} model. The fact that \emph{WML4} gives a result consistent with that for the \emph{MILL} model, suggests that the exact value of $\sigma_8$ does not change the measured value of the halo mass significantly. 

This result also suggests that the fact that we compared the $z=2.25$ simulation snapshot with observations at $z=2.36$ should not have any significant effect on our result. The difference in redshift corresponds to a slight change in $\sigma_8$, and we have shown that the exact value of this parameter does not change the results significantly.

For comparison, we show 2-D Ly$\alpha$ absorption maps for different models in Figure~\ref{models}, for haloes with $M_{\rm h}>10^{11.5}\Msun$. Model \emph{AGN} shows weaker absorption within $\sim300$ kpc from galaxies in comparison with the other models and \emph{WML4} shows less absorption relative to the \emph{REF} model.
\figbb

The implied minimum masses from all models are summarized in Table~\ref{results.tab} and Figure~\ref{masscomparison}. All models are consistent with each other at the 1$\sigma$ level. It is thus clear that the statistical errors on the minimum halo mass, which are around 0.2-0.3 dex, are comparable to or larger than the systematic errors arising due to uncertainties in the physics important for galaxy formation, as inferred from simulations with different feedback prescriptions and different cosmologies. This demonstrates the robustness of this method for measuring halo masses. We also list the reduced $\chi^{2}$ values between the observed map and the best matching simulated maps in the considered region, i.e. for  impact parameters 0-2 pMpc and velocity separations $154-616\,\rm km\, s^{-1}$, with the caveat that the distance bins are correlated along the LOS on the $\sim1$ pMpc scale (R12).

	\begin{table}
    	\begin{minipage}{\linewidth}
      	\caption{The minimum halo masses (with 1 and $2\sigma$ confidence intervals) of the $\langle z \rangle \approx2.36$ star-forming galaxies of R12 inferred from a comparison of the observed \ion{H}{I} absorption around them to predictions from different hydrodynamical simulations. Only impact parameters 0-2 pMpc and velocity separations $154-616\,\rm km\, s^{-1}$ are considered. The corresponding number densities of haloes (and $1$ and $2\sigma$ confidence intervals) as measured from the simulations with $100\, h^{-1}\rm\, cMpc$ boxes are presented in the third  column. For reference, the observed number density of galaxies in the photometric sample, from  which the objects in R12 were drawn for spectroscopic follow-up, is $11\times10^{-3}\, h^{3}\, \rm cMpc^{-3}$, with an uncertainty of 10\% \citep{Reddy2008}. The last column shows the reduced $\chi^{2}$ between the observed map and the best matching simulated map, for  impact parameters 0-2 pMpc and velocity separations $154-616\,\rm km\, s^{-1}$.}
      	\begin{tabular}{l|lll|}

        	\hline
        	\noalign{\smallskip}
	Models &Min. halo mass [log$_{10}\Msun$] & n [$10^{-3} h^{3}\rm\,cMpc^{-3}$] & Red. $\chi^{2}$ \\ 
	\noalign{\smallskip}
	\hline
	\noalign{\smallskip}
	\emph{REF}  	&	${11.6}^{+0.20+0.50}_{-0.20-0.60}$ 	&   	$2.9^{+2.2+12.6}_{-1.2-2.2}$ & 1.018 \\
	\emph{AGN}	&	${11.9}^{+0.20+0.30}_{-0.30-0.70}$ & 	$1.0^{+1.6+7.3}_{-0.5-0.7} $   & 0.934 \\	
	\emph{MILL} 	& 	${11.7}^{+0.05+0.30}_{-0.40-0.90}$ 	& 	$4.1^{+6.7+33.3}_{-0.5-2.1}$ & 0.989 \\
	\emph{WML4}	& 	${11.4}^{+0.50+0.80}_{-0.10-0.10}$ 	& 	$4.7^{+1.5+1.5}_{-3.5-4.3} $  & 1.091 \\
	
        	\noalign{\smallskip}
	\hline
      	\end{tabular}
      	\label{results.tab}
    	\end{minipage}
  	\end{table}
	
\figcb

\subsubsection{Number densities of haloes}
The number density of galaxies in the photometric sample, from which the galaxies in R12 were drawn, is $11\times10^{-3}\, h^{3}\, \rm cMpc^{-3}$, with an uncertainty of 10\% \citep{Reddy2008}. The largest weights for spectroscopic follow-up were given to objects in the apparent magnitude range $R=23-24.5$ \citep{Steidel2004}, while the limit for the photometric sample is $R=25.5$. In addition, objects near the QSO sightlines (i.e.\  within 1-2 arcminutes) were given extra weight. There are $\sim2.5$ times more objects in the photometric sample for the limiting magnitude of $R=25.5$ than for $R=24.5$, and so the lower limit on the number density of galaxies in the spectroscopic sample is $\sim4.5\times10^{-3}\, h^{3}\, \rm cMpc^{-3}$. The real number density of objects in the spectroscopic sample is somewhere in between these numbers, because a non-negligible fraction of objects with $R>24.5$ were followed up spectroscopically.

Table~\ref{results.tab} lists the number densities corresponding to the minimum halo masses inferred for each model. Their values are  $2.9^{+2.2+12.6}_{-1.2-2.2}\times10^{-3}$, $1.0^{+1.6+7.3}_{-0.5-0.7}\times10^{-3}$, $4.1^{+6.7+33.3}_{-0.5-2.1}\times10^{-3}$, and $4.7^{+1.5+1.5}_{-3.5-4.3}\times10^{-3}\, h^{3}\, \rm cMpc^{-3}$ (the first and second pairs of errors indicate 1$\sigma$ and 2$\sigma$ confidence intervals, respectively) for models \emph{REF}, \emph{AGN}, \emph{MILL} and \emph{WML4}, respectively. All the estimates, apart from \emph{AGN}, are consistent with the observed number densities within 1$\sigma$, and \emph{AGN} agrees with the observations within 2$\sigma$. The minimum halo mass for the \emph{MILL} model is 0.1 dex higher than that for the \emph{REF} model, and yet the number density is also somewhat higher. This is due to the higher value of $\sigma_8$ in the \emph{MILL} model (0.9 as compared to 0.74 for the other models), which causes faster growth of structure (see Figure~\ref{medianhalomass}). Although the inferred number density is lower for the \emph{AGN} model than for the other models, they are consistent at the 1$\sigma$ level due to the uncertainty on the minimum halo mass estimate.

\subsubsection{Comparison with estimates from clustering measurements}
Estimates of halo masses of $z\sim2$ star-forming galaxies were recently presented by \citet{Conroy2008} and Trainor et al. (2012). As already discussed in the introduction, using a clustering analysis \citet{Conroy2008} found the minimum halo mass to be $10^{11.55\pm0.2}\Msun$ for the WMAP year-1 cosmology, and  0.3-0.4 dex smaller (i.e.\ $10^{11.15}-10^{11.25}\Msun$) for the WMAP year-3 cosmology (which is the default cosmology in our study). \citet{Trainor2012} measure masses of $M_{\rm h}>10^{11.7}\Msun$ for the spectroscopic sample using WMAP year-7 cosmological parameters (where $\sigma_{8}$  is intermediate between those from WMAP year-1 and year-3, while the value for $\Omega_{\rm m}$ is closer to that from WMAP year-1).  Our results for the models that use WMAP year-1 and year-3 cosmological parameters (log$_{10}M_{\rm min}/\Msun={11.7}^{+0.05+0.30}_{-0.40-0.90}$ for the \emph{MILL} model which uses WMAP1, and ${11.4}^{+0.50+0.80}_{-0.10-0.10}$ for the \emph{WML4} model which uses WMAP3), are in good agreement with these estimates. We conclude that the two methods (i.e.\ clustering analysis and matching of the Ly$\alpha$ absorption profiles) give consistent results.

\subsubsection{Measuring mass from pencil beam surveys}

We have demonstrated that comparisons of observed 2-D Ly$\alpha$ absorption maps around galaxies with maps extracted from cosmological, hydrodynamical simulations can be used to measure the halo masses of the galaxy population in question. Our observed QSO-galaxy  fields typically span areas of $\approx5\times7$ arcminutes with the QSO in the middle. Here we test the performance of the method for a limited  survey area of only $\approx30\times30$ arcseconds, which corresponds to a cut along the LOS through the first impact parameter bin of the above 2-D maps ($b\lesssim130$ pkpc and velocity separation $154<v_{\rm LOS}<616\,\rm km\, s^{-1}$). The implied minimum halo mass is log$_{10}M_{\rm min}/\Msun={11.6}^{+0.40+0.50}_{-0.05-0.70}$, which agrees very well with that inferred from a comparison of the full maps (log$_{10}M_{\rm min}/\Msun={11.6}^{+0.20+0.50}_{-0.20-0.60}$), also in terms of the size of the error bars. 

This finding is relevant for future galaxy-QSO pair surveys such as the one planned with the VLT/MUSE integral field spectrograph \citep{Bacon2010}, whose field of view is $1\times1$ arcminute. Applying this method to MUSE observations will allow estimates of halo masses of Ly$\alpha$ emitters from a single telescope pointing on each QSO in the survey. This eliminates the need to make a wide field mosaic to measure clustering, which would be too expensive for the faint objects dominating the number density of galaxies detected in the deep fields.

\section{Redshift space anisotropies}\label{zdistortions}

R12 reported two types of redshift space anisotropies in the observed 2-D Ly$\alpha$ absorption maps. At small impact parameters ($\lesssim200$ pkpc) the absorption signal is elongated along the LOS, which is likely a result of gas peculiar velocities of $\approx200\, \rm km\, s^{-1}$, and errors in galaxy redshifts ($\approx125\,\rm km\, s^{-1}$). On the other hand, at large separations from galaxies ($\approx1.5$ pMpc) the absorption appears compressed along the LOS, which R12 attributed to large-scale gas infall into the potential wells populated by the observed galaxies. 

Here we use simulated observations to examine the origin of the observed redshift space anisotropies. Figure \ref{2D_2} shows 2-D absorption maps centered on haloes with masses higher than the minimum mass inferred in the previous section, i.e.\  $\rm M_{\rm h}>10^{11.5}\Msun$, for 4 cases: \emph{i)} the default case, i.e.\  redshift errors with $\sigma=60\rm\, km\, s^{-1}$ for 10$\%$ of galaxy redshifts, and $\sigma=125\rm\, km\, s^{-1}$ for the remaining 90$\%$ of redshifts, and taking into account peculiar velocities, \emph{ii)} no redshift errors, \emph{iii)} ignoring peculiar velocities, and \emph{iv}) a number of galaxies per impact parameter bin identical to the observations of R12. 
\figd

Comparing the 2nd and 3rd panels of Figure~\ref{2D_2}, we see that the effect of redshift errors is to wash out the signal along the LOS. 

The 4th panel shows the case where we ignore peculiar velocities. At small impact parameters ($\lesssim200$ pkpc), the absorption is much reduced, demonstrating that peculiar velocities strongly enhance the small-scale absorption excess. In addition, on large scales ($\gtrsim1.5$ pMpc) the absorption enhancement is now much more isotropic, demonstrating that the observed large-scale ($\gtrsim1.5$ Mpc) compression along the LOS is likely due to large-scale infall. The observations of R12 are the first of gas infall into large-scale potential wells. We just note here that, as expected, ignoring both peculiar velocities and redshift errors results in a map with axisymmetric absorption (not shown). 

In the 5th panel we use the same number of sightlines per impact parameter bin as in the observations. The resulting map is noisier in comparison with the previous panels with simulated maps, but the redshift space distortions seen in the first panel are also clearly visible in this panel. 

A more quantitative comparison is provided by Figures~\ref{2D_LOScuts_2}  and~\ref{2D_TDcuts_2} which show cuts through these maps along the LOS and in the transverse direction, respectively, for three different cases. The top left panel of Figure~\ref{2D_LOScuts_2} makes it very clear that the simulations  underpredict the amount of absorption for impact parameters smaller than $\approx100$ pkpc for velocities $\lesssim154\,\rm km\, s^{-1}$. Conversely, at impact parameters 0.13-0.25 pMpc the simulations overestimate the absorption for velocities $\lesssim154\,\rm km\, s^{-1}$. However, the number of observed galaxies in these two impact parameter bins is too small (15 and 8, respectively) to obtain an accurate estimate of the significance of these discrepancies. Note also that a simple ``$\chi$ by eye" would strongly overestimate the significance of these discrepancies because the points are strongly correlated in the LOS direction (R12). 

Both figures show that redshift errors smooth the signal along the LOS on scales $\lesssim10^2\rm\, km\, s^{-1}$. The orange dashed curves, which do not take peculiar velocities into account, lie below the default case (solid black curves) for velocity separations $\lesssim150\rm\, km\, s^{-1}$, and above them at larger velocity separations. The increased optical depth closer to galaxies could be due to large-scale infall of gas enhancing absorption through ``filling'' in redshift space. Another possibility is that the enhanced absorption is due to lines getting broadened due to large-scale galactic outflows, which could also significantly affect our median statistics. 

Comparing the bottom-right panels of Figures~\ref{2D_LOScuts_2}  and~\ref{2D_TDcuts_2}, which show the absorption for large impact parameters and large velocity separations, respectively, we see that the absorption is observed to be stronger along the LOS (i.e.\ in Figure~\ref{2D_LOScuts_2}) than transverse to the LOS (i.e.\ in Figure~\ref{2D_TDcuts_2}). The simulations show the same anisotropy (compare solid black curves in the bottom-right panels of Figures~\ref{2D_LOScuts_2}  and~\ref{2D_TDcuts_2}), and the anisotropy is enhanced when redshift errors are ignored (dotted blue curves). In contrast, when peculiar velocities are ignored (dashed orange curves) the LOS and transverse directions are consistent with each other. Thus, the simulations support the interpretation of R12 of the large-scale anisotropy as being due to infall.
 
\fige
\figf

In Figure~\ref{2D_zerrors} we add different amounts of scatter to the galaxy redshifts. Large redshift errors smooth the absorption signal more along the LOS. The last panel shows the case with redshift errors  of $\sigma=300\rm\, km\, s^{-1}$, where instead of compression along the LOS relative to the direction transverse to the LOS, we see elongation. This suggests that the errors in galaxy redshifts have to be $<200\rm\, km\, s^{-1}$ to study large-scale infall of gas into the potential wells of star-forming galaxies.

\figda

\section{Summary \& Conclusions}\label{Summary}
Observations have shown that the absorption of the light from background QSOs is enhanced at the rest-frame wavelength of \ion{H}{I} Ly$\alpha$ when a sightline passes within  several pMpc from a galaxy. We compared the observed median Ly$\alpha$ absorption profiles around star-forming galaxies at $z\approx2.36$ from \citet{Rakic2012}  with predictions from cosmological SPH simulations in order to statistically constrain the total masses of the haloes hosting the observed galaxy population, and to explain the observed redshift space distortions. We confined our comparison to transverse separations 0-2 pMpc and to LOS velocity differences $154-616\,\rm km\, s^{-1}$. The large-scale cut avoids comparing regions without signal and the small-scale cut excludes the region most affected by redshift errors and uncertain physical processes.

The most important conclusions of this study are:
\begin{itemize}
\item Using the \emph{REF} model from the \textsc{OWLS} suite of simulations, we derive a minimum halo mass of  log$_{10}M_{\rm min}/\Msun={11.6}^{+0.20+0.50}_{-0.20-0.60}$, where the first and second pairs of errors indicate the 1 and 2$\sigma$ uncertainties, respectively. This is in good agreement with the estimates from the clustering analyses of \citet{Conroy2008} and Trainor et al. (2012) performed on the same galaxy population. 
\item This method is robust to changes in feedback prescriptions as demonstrated by comparison with the \emph{AGN} model in which galaxies eject large amounts of gas at the redshift of interest here. The implied minimum mass of log$_{10}M_{\rm min}/\Msun={11.9}^{+0.20+0.30}_{-0.30-0.70}$, although 0.3 dex higher, is consistent with the \emph{REF} model within 1$\sigma$.
\item The minimum halo mass inferred from a comparison with the \emph{MILL} simulation (log$_{10}M_{\rm min}/\Msun={11.7}^{+0.05+0.30}_{-0.40-0.90}$) is consistent with that required by the \emph{WML4} model (log$_{10}M_{\rm min}/\Msun={11.4}^{+0.50+0.80}_{-0.10-0.10}$). As these two models assume different cosmologies (WMAP year-1 vs. WMAP year-3), but are otherwise identical, this suggests that the method is not very sensitive to differences in cosmological parameters.
\item If we consider only impact parameters $\lesssim100$ pkpc and LOS velocity differences $154-616\,\rm km\, s^{-1}$, then we infer a mass of log$_{10}M_{\rm min}/\Msun={11.6}^{+0.40+0.50}_{-0.05-0.70}$, which is almost identical to the one obtained from the full 2-D map (impact parameters $<2$ pMpc). This is very encouraging for future narrow field QSO-galaxy surveys (e.g.\ the one planned with the VLT/MUSE), where one will be able to estimate halo masses of Ly$\alpha$ emitters with a single telescope pointing, without the need for a wide field mosaic. 
\item The observed elongation of the absorption signal of $\approx200\,\rm km\, s^{-1}$ at small impact parameters is a product of uncertainties in galaxy redshifts, observed to be $\approx125\,\rm km\, s^{-1}$, and the peculiar motions of gas in and around galaxy haloes. 
\item At impact parameters $<0.13$ pMpc the simulations may under-predict the absorption for velocities $<154\,\rm km\, s^{-1}$. Conversely, at impact parameters 0.13-0.25 pMpc the simulations may over-predict the absorption in the same velocity range. However, the number of observed galaxies in these two impact parameter bins is too small (15 and 8, respectively) to estimate the significance of these discrepancies. 
\item The compression of the signal on large scales ($\gtrsim1.5$ pMpc) is a result of gas infall into the potential wells occupied by the galaxies. To observe this infall, the accuracy of galaxy redshifts has to be significantly better than $200\,\rm km\, s^{-1}$, as larger redshift errors smooth the absorption signal along the LOS, disguising the compression signature.
\end{itemize}

\section*{Acknowledgments}
The authors would like to thank all the members of the OWLS team for useful discussions and the anonymous referee whose suggestions have improved the clarity of the paper. The simulations presented here were
run on Stella, the LOFAR BlueGene/L system in Groningen, on
the Cosmology Machine at the Institute for Computational Cosmology
in Durham (which is part of the DiRAC Facility jointly
funded by STFC, the Large Facilities Capital Fund of BIS, and
Durham University) as part of the Virgo Consortium research programme,
and on Darwin in Cambridge. This work was sponsored
by the National Computing Facilities Foundation (NCF) for the use
of supercomputer facilities, with financial support from the Netherlands
Organization for Scientific Research (NWO), also through
a VIDI grant. The research leading to these results has received
funding from the European Research Council under the European
Unions Seventh Framework Programme (FP7/2007-2013) / ERC
Grant agreement 278594-GasAroundGalaxies and from the Marie
Curie Training Network CosmoComp (PITN-GA-2009-238356).

\appendix

\section{Convergence Tests}\label{App:ResolutionTests}
\begin{figure*}

\quad
\includegraphics[width=0.33\textwidth]{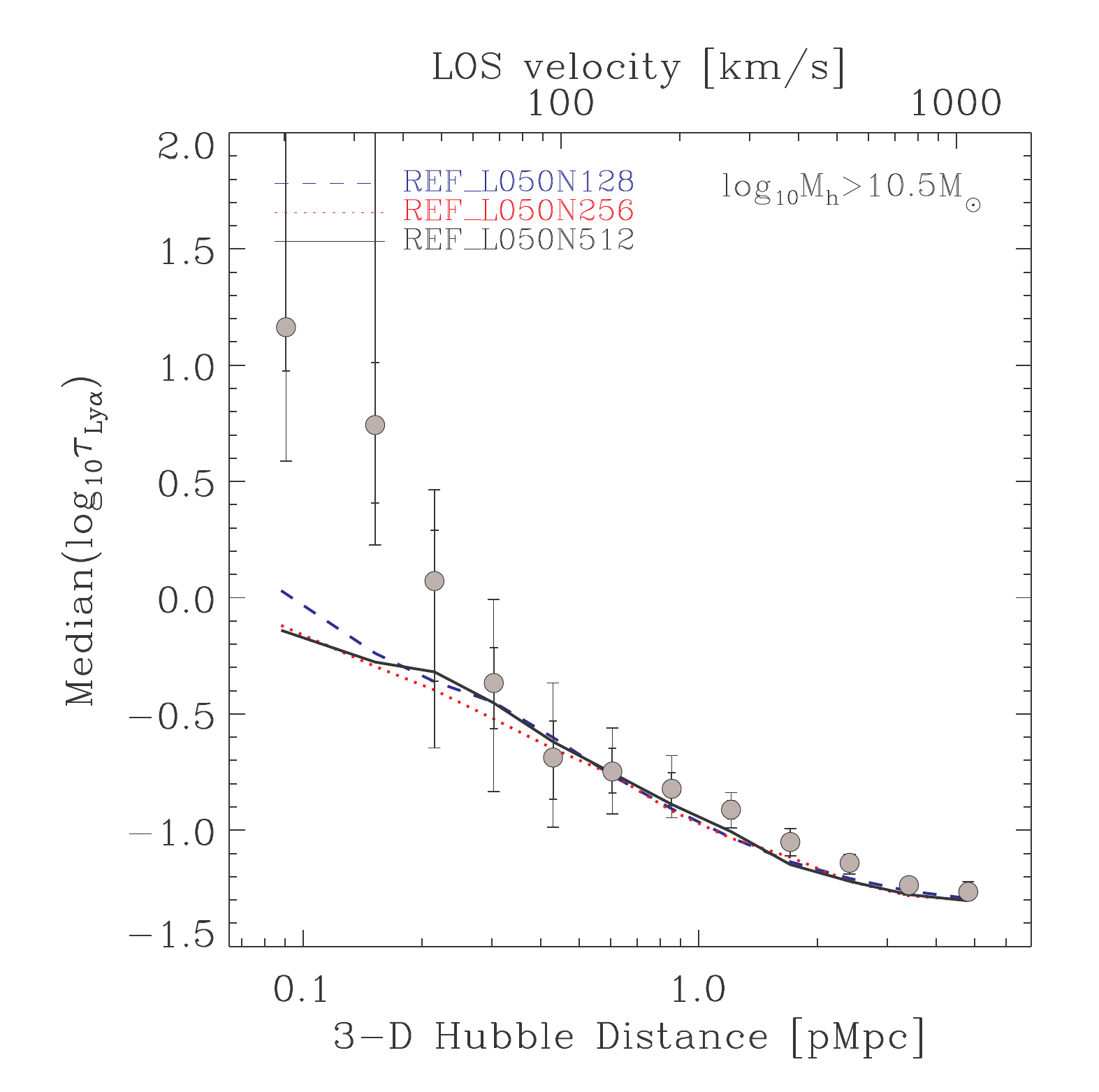}\includegraphics[width=0.33\textwidth]{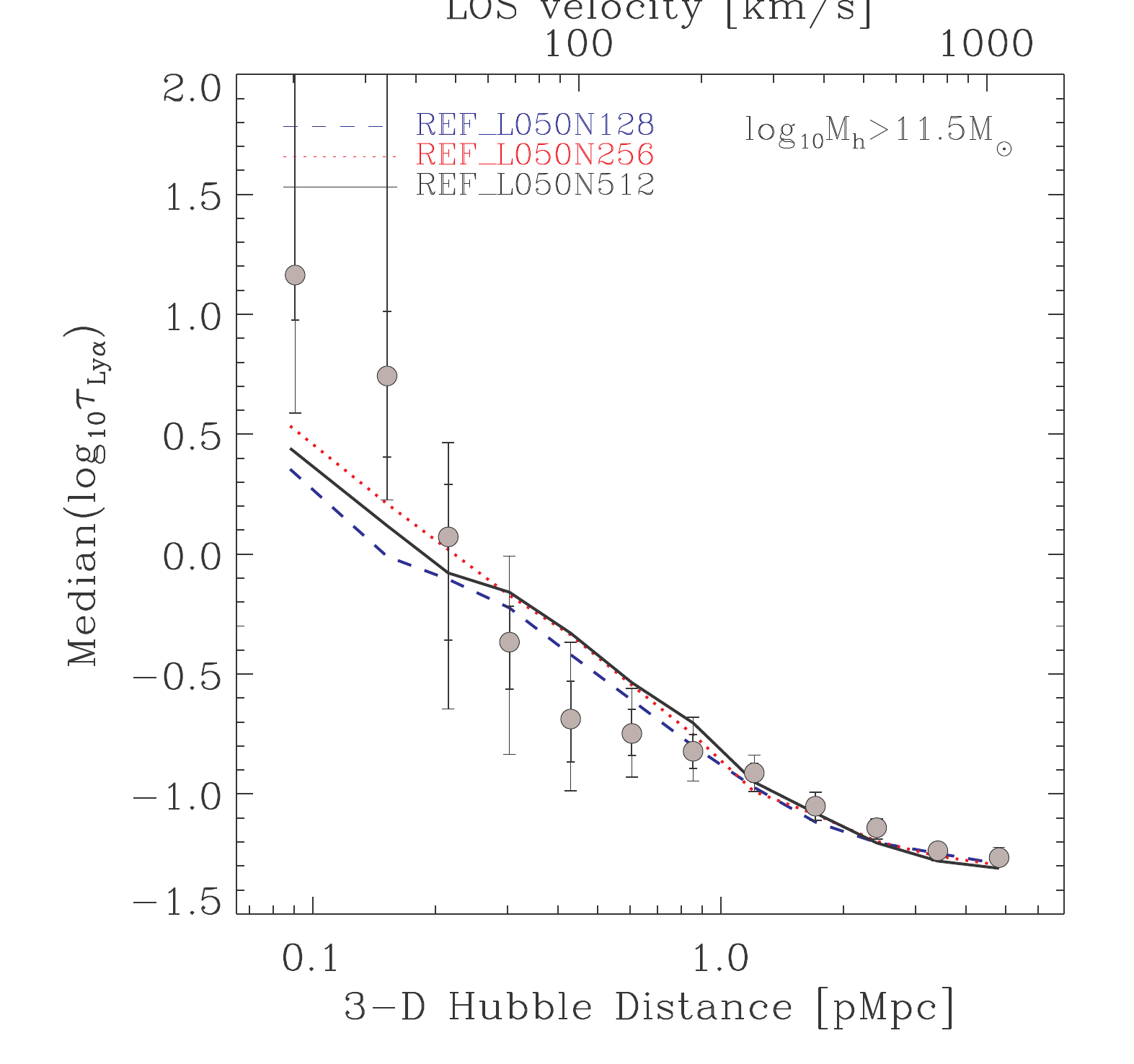}\includegraphics[width=0.33\textwidth]{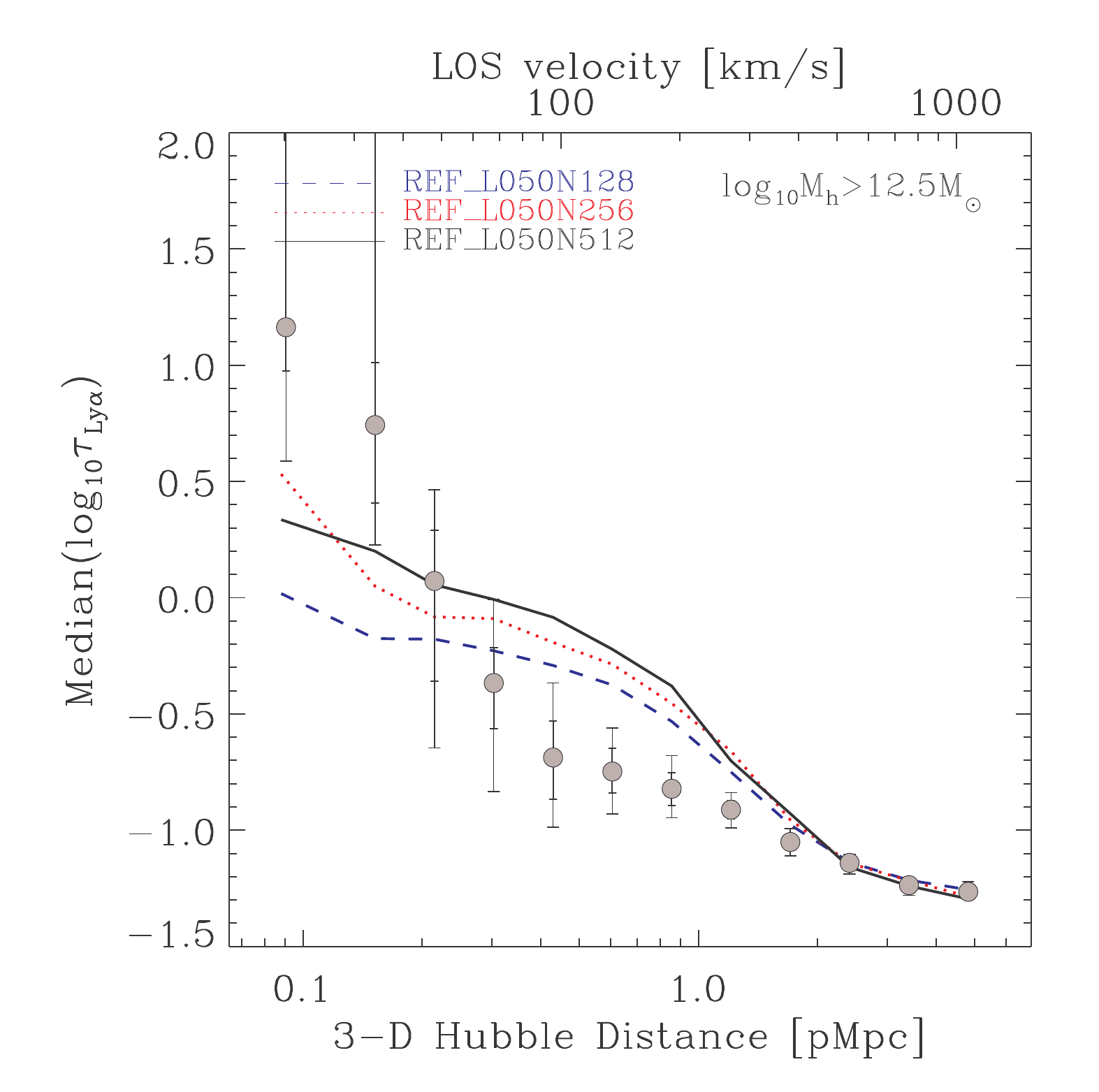}
\caption{Convergence with respect to the numerical resolution: median log$_{10}(\tau_{\rm Ly\alpha}$) as a function of proper 3-D Hubble distance from galaxy haloes in the 50 $h^{-1}$ cMpc \emph{REF} simulations, with $128^3$, $256^3$, and $512^3$ particles. Each panel shows the results for a different halo mass range, as indicated in the panels, while grey symbols with $1\sigma$ and 2$\sigma$ error bars  show the observations of R12. We added  errors to the galaxy positions in simulations to mimic redshift errors (see text for more details).   \label{HubbleRes}}  
\end{figure*}

Figure~\ref{HubbleRes} shows a resolution test for the 50 $h^{-1}$ cMpc \emph{REF} simulation, performed with $128^3$, $256^3$, and $512^3$ particles. The median log$_{10}(\tau_{\rm Ly\alpha}$) curves  as a function of 3-D proper Hubble distance from the galaxies are  consistent with each other  for all but the highest mass bin. In the highest mass bin (right panel) the resulting curve for the simulation with the poorest resolution is below the curves for the simulations with higher resolution, but the simulations with $256^3$, and $512^3$ particles give consistent results. Therefore, we conclude that the simulation results are converged with respect to the numerical resolution. 

\begin{figure*}

\quad
\includegraphics[width=0.33\textwidth]{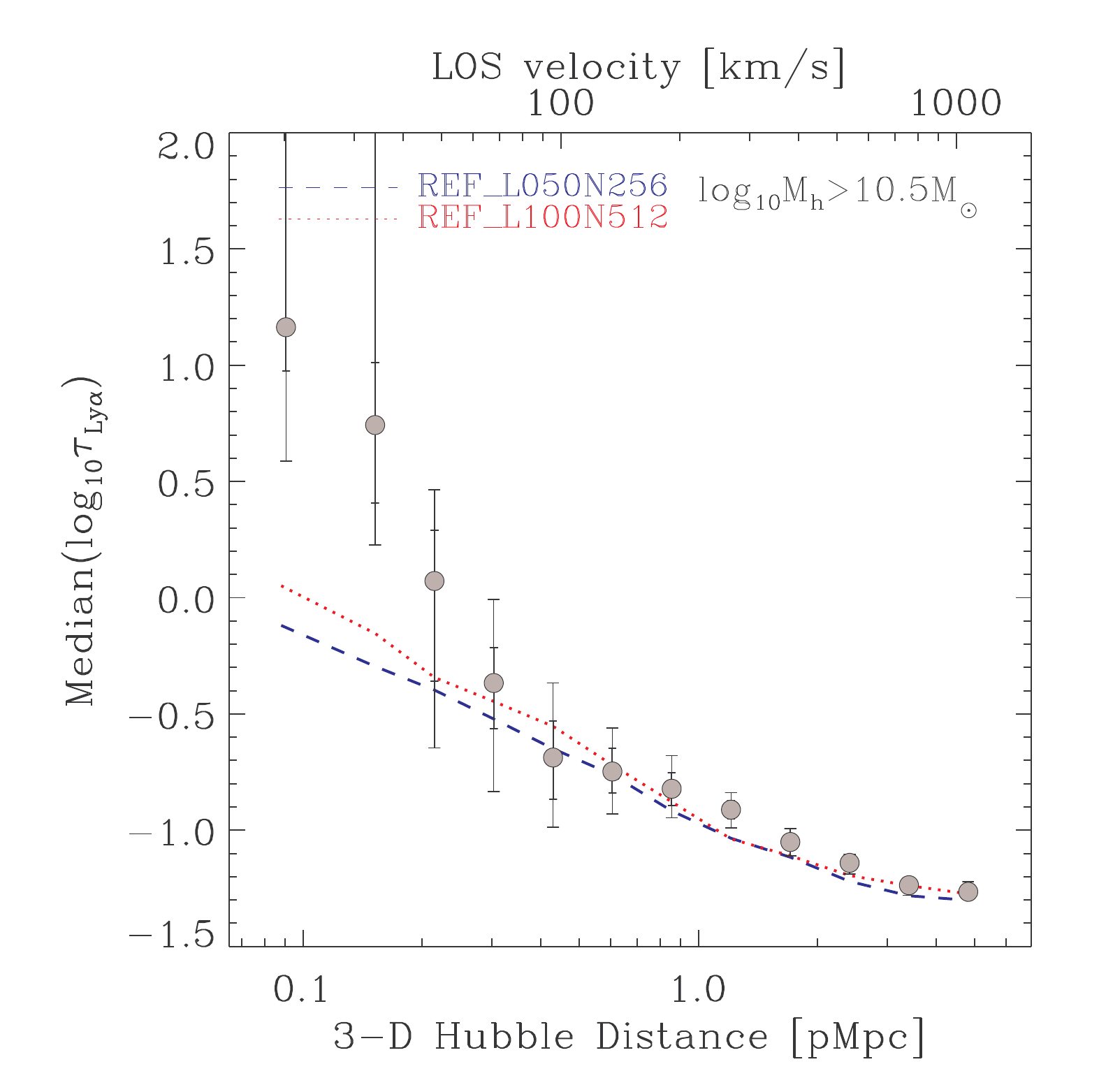}\includegraphics[width=0.33\textwidth]{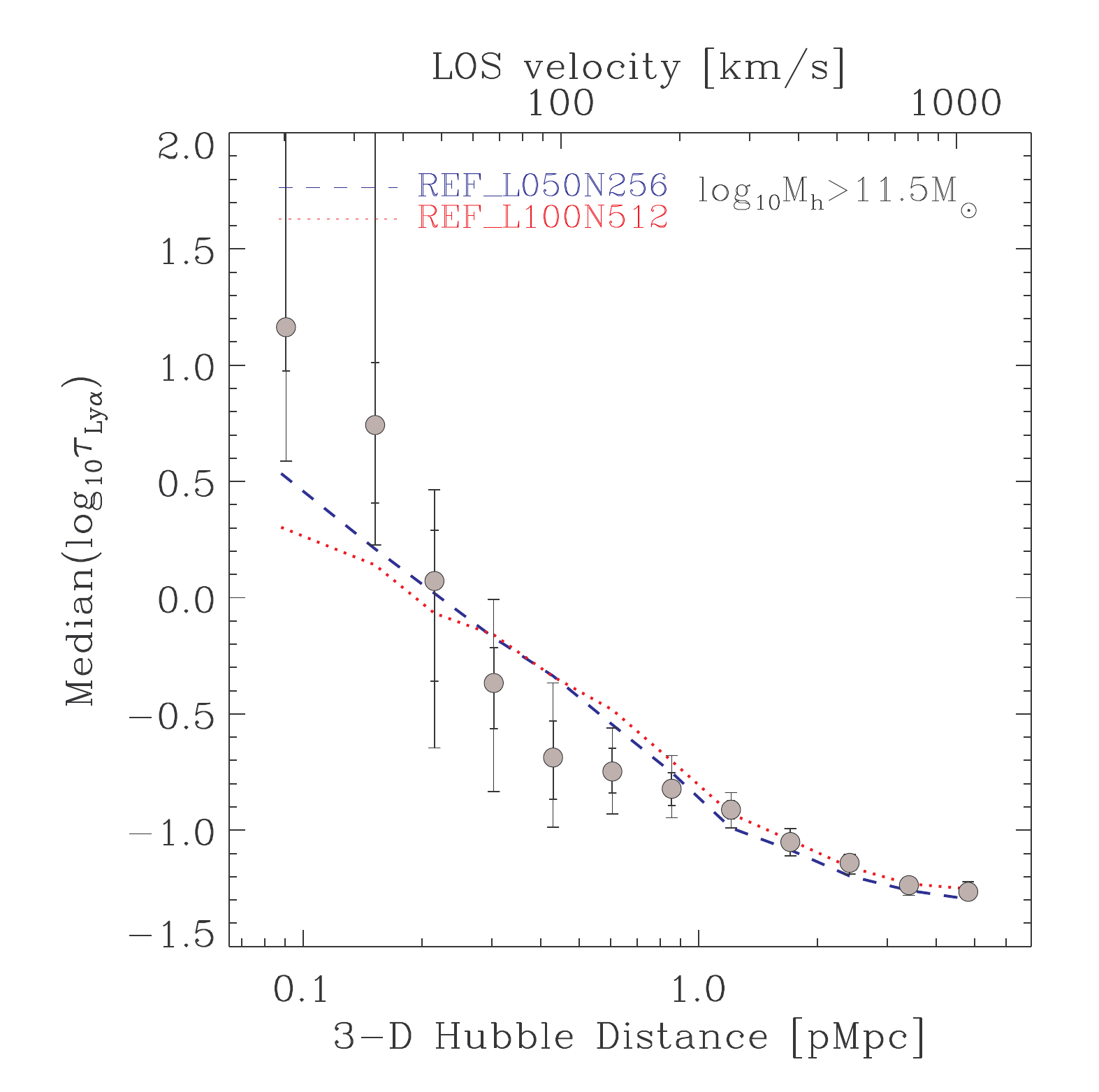}\includegraphics[width=0.33\textwidth]{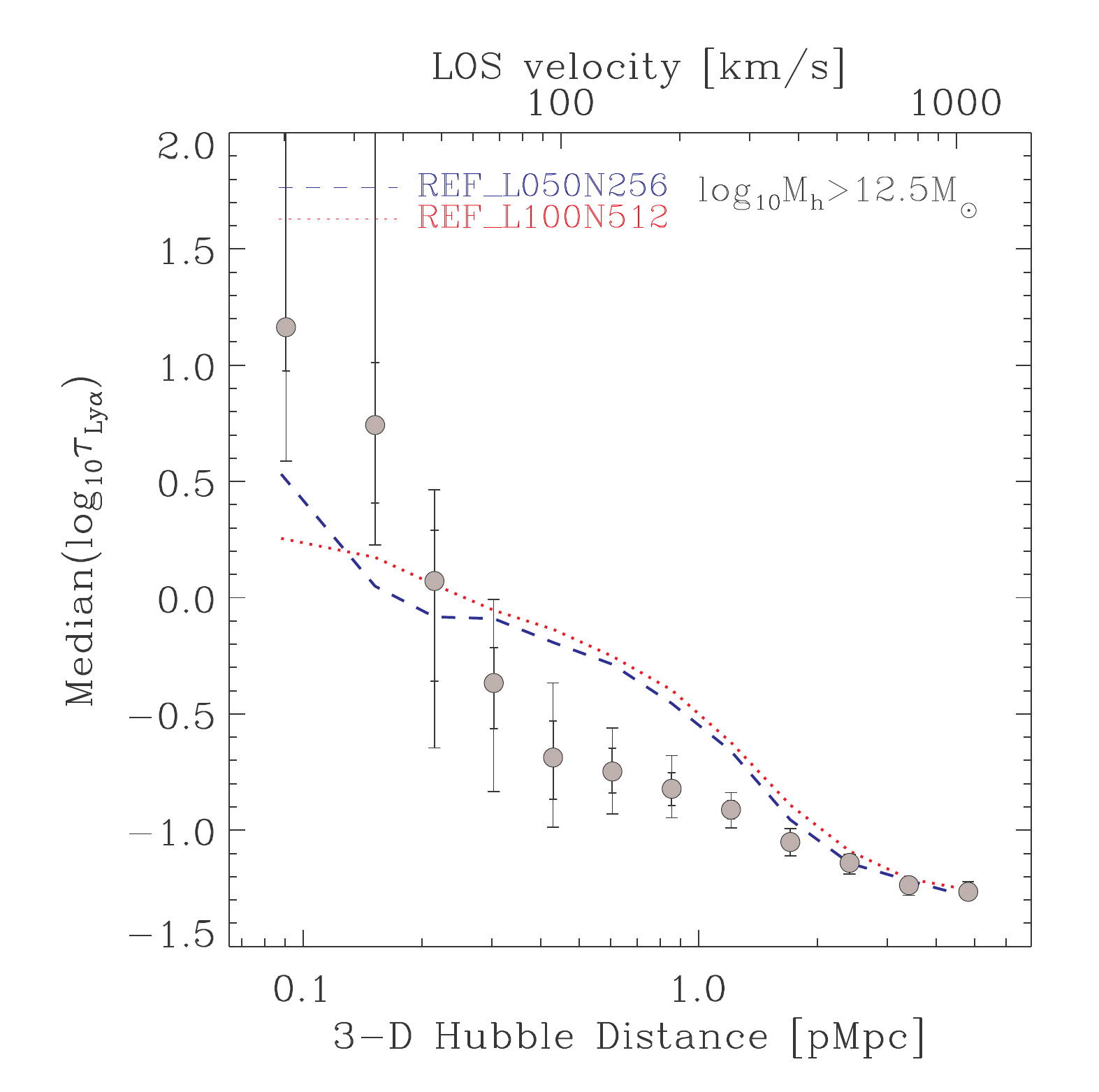}
\caption{Convergence with respect to the size of the simulation box: Median log$_{10}(\tau_{\rm Ly\alpha}$) as a function of proper 3-D Hubble distance from galaxy haloes in the 50 $h^{-1}$ cMpc \emph{REF} simulation with $256^3$, and in the 100 $h^{-1}$ cMpc \emph{REF} simulation with $512^3$ particles. Each panel shows the results for a different halo mass range, as indicated in the panels, while grey symbols with $1\sigma$ and 2$\sigma$ error bars  show the results from the observations of R12. We added  errors to the galaxy positions in simulations to mimic redshift errors (see text for more details).   \label{HubbleBox}}  
\end{figure*}

Figure~\ref{HubbleBox} compares the results for  the 50 $h^{-1}$ cMpc \emph{REF} simulation with $256^3$, and the 100 $h^{-1}$ cMpc \emph{REF} simulation with $512^3$ particles. In this case the particle masses in the two simulations are the same, but the simulation volume is changed. The median optical depth curves show that the results are converged with respect to the box size.

\bsp

\label{lastpage}

\end{document}